\documentclass[useAMS,usenatbib]{mn2e}

\usepackage{color}
\usepackage{multirow}         %to use 
\usepackage{graphicx,epsfig}
\usepackage{multicol}
\def \zbar {\bar{\rm z}}
\def \Mpc {{\rm Mpc} }

\def \hmpc{~h^{-1}\,{\rm Mpc}} 
\def \hkpc{~h^{-1}\,{\rm kpc}}

\def \gsim { \lower .75ex \hbox{$\sim$} \llap{\raise .27ex \hbox{$>$}} }
\def \lsim { \lower .75ex \hbox{$\sim$} \llap{\raise .27ex \hbox{$<$}} }
\def \deg {^{\circ}}

\def \hmsun {~h^{-1}\,M_{\odot}}

\def\be{\begin{equation}}
\def\ee{\end{equation}}
\title[Clustering evolution and a search for BAO]{Angular correlation
function of 1.5 million LRGs: clustering evolution and a search for BAO}
\author[U. Sawangwit et al.]{U. Sawangwit$^1$\thanks{E-mail:
utane.sawangwit@durham.ac.uk}, T. Shanks$^1$, F.\,B. Abdalla$^2$, R.\,D. Cannon$^3$, S.\,M. Croom$^4$, 
\newauthor A.\,C. Edge$^5$, Nicholas P. Ross$^{1,6}$ and D.\,A. Wake$^{1,7}$\\
$^1$Physics Department, University of Durham, South Road, Durham, DH1 3LE, UK\\
$^2$Department of Physics and Astronomy, University College London, Gower Street, London, WC1E 6BT, UK\\
$^3$Anglo-Australian Observatory, PO Box 296, Epping, NSW 1710, Australia\\
$^4$School of Physics, University of Sydney, NSW 2006, Australia\\
$^5$Institute for Computational Cosmology, University of Durham, South Road, Durham, DH1 3LE, UK\\
$^6$Department of Astronomy and Astrophysics, The Pennsylvania State University, University Park, PA 16802, USA\\
$^7$Department of Astronomy, Yale University, CT 06520, USA}

\begin{document}
\date{Accepted 2011 June 14. Received 2011 May 24; in original form 2009 November 25}
%\date{Accepted ????  Received ????; in original form ????}
\pagerange{\pageref{firstpage}--\pageref{lastpage}} \pubyear{2011}

\maketitle
\label{firstpage}
\begin{abstract}
We present the angular correlation function measured from photometric samples
comprising 1\,562\,800 luminous red galaxies (LRGs). Three LRG samples were
extracted from the Sloan Digital Sky Survey (SDSS) imaging data, based on
colour-cut selections at redshifts, $z \approx 0.35, 0.55 ~\rm{and} ~0.7$ as
calibrated by the spectroscopic surveys, SDSS-LRG, 2dF-SDSS LRG and QSO (2SLAQ),
and the AAOmega LRG survey. The galaxy samples cover $\approx 7600$ deg$^2$ of
sky, probing a total cosmic volume of $\approx 5.5 ~h^{-3}\,\rm{Gpc}^{3}$.

The small and intermediate scale correlation functions generally show
significant deviations from a single power-law fit with a well-detected break at
$\approx 1 \hmpc$, consistent with the transition scale between the 1-- and 2--
halo terms in halo occupation models. For galaxy separations $1-20\hmpc$ and at
fixed luminosity, we see virtually no evolution of the clustering with redshift
and the data is consistent with a simple high peaks biasing model where the
comoving LRG space density is constant with $z$. At fixed $z$, the LRG
clustering amplitude increases with luminosity in accordance with the simple
high peaks model, with a typical LRG dark matter halo mass $10^{13}-10^{14}\hmsun$.
For $r < 1 \hmpc$, the evolution is slightly faster and the clustering decreases towards
high redshift consistent with a virialised clustering model. 
However, assuming the HOD and $\Lambda$CDM halo merger frameworks, $\sim 2-3$ per\,cent/Gyr 
of the LRGs is required to merge in order to explain the small scales clustering evolution, 
consistent with previous results. 

At large scales, our result shows good agreement with the SDSS LRG result of 
Eisenstein et al. (2005) but we find an apparent excess clustering signal beyond 
the BAO scale. Angular power spectrum analyses of similar LRG samples also detect 
a similar apparent large-scale clustering excess but more data is required to check 
for this feature in independent galaxy datasets. Certainly, if the 
$\Lambda$CDM model were correct then we would have to conclude that this excess 
was caused by systematics at the level of $\Delta w\approx0.001-0.0015$ in the 
photometric AAOmega-LRG sample.

\end{abstract}

%\nokeywords
%\begin{keywords}
%\end{keywords}
\section{Introduction}
\label{sec:intro}

The galaxy two-point function whether in its correlation function or power
spectrum form has long been recognised as a powerful statistical tool for studying
Large-Scale Structure (LSS) of the Universe \citep{Peebles80}. In an isotropic
and homogeneous Universe, if the density fluctuation arises from a Gaussian
random process, the two-point correlation function, $\xi(r)$, and its Fourier
transform, $P(k)$, contains a complete description of such fluctuations. It has
been used to measure the clustering strength of galaxies in numerous galaxy
surveys (see e.g. \citealt{Groth77,Shanks89,Baugh93,Rat98}) and the observed
$\xi(r)$ is reasonably well represented by a power-law of the form
$\xi(r)=(r/r_{0})^{-1.8}$ over a large range of scales, $\approx100 \hkpc$ --
$10 \hmpc$, where $r_{0}$ is approximately $5 \hmpc$.

More recently, large galaxy redshift surveys have become available
(SDSS:\citealt{York00}, 2dFGRS:\citealt{Colless01}) and these surveys provide a
perfect opportunity to exploit the two-point function as a tool to constrain
cosmological parameters
\citep{Hawkins03,Cole05,Eisenstein05,Tegmark06,Percival07} which in turn
provides an excellent test for our current understanding of the Universe and the
processes by which the LSS were formed. 

In the past, when galaxy redshift surveys were less available, the angular
correlation function, $w(\theta)$, was heavily utilised in the analysis of
imaging galaxy samples. The spatial correlation function, $\xi(r)$, can be
related to $w(\theta)$ via Limber's equation \citep{Limber53}, alternatively
$w(\theta)$ can be inverted to $\xi(r)$ using Lucy's iterative technique
(\citealt{Lucy1974}), both approaches providing a means to recover the 3--D
clustering information numerically. Even today, galaxy imaging 
surveys still tend to cover a bigger area of the sky and occupy a larger volume
than redshift surveys and therefore could offer a route to a more
accurate estimation of the correlation function and power spectrum (see e.g.
\citealt{Baugh93}). One of the disavantages of using $w(\theta)$ is the dilution 
of the clustering signal from projection and hence any small-scale/sharp feature 
which might exist in the 3--D clustering may not be observable in $w(\theta)$.

As mentioned above, the correlation function at small to intermediate scales can
be approximately described by a single power-law which also results in a
power-law $w(\theta)$ but with a slope of $1-\gamma$. However with larger sample
sizes, recent analyses of galaxy distributions start to show a deviation from a
simple power-law (\citealt{Zehavi05b,Phleps06,Ross07,Blake08}, see also
\citealt{Shanks83}). This poses a challenge for a physical explanation and
understanding of non-linear evolution of structure formation. Several authors
attempted to fit such correlation function using a description of halo model
framework \citep[e.g.][]{Cooray02} invoking a transition between 1-- and 2--
halo terms which occurs at $\approx 1 \hmpc$ where the feature is observed. 
This distance scale could potentially be used as a `standard ruler' in tracking
the expansion history of the Universe, provided that its physical origin is well
understood and the scale can be accurately calibrated.

Another feature in the correlation function predicted by the standard 
$\Lambda$CDM model is the `Baryon Acoustic Oscillations' (BAO). BAO arise from
sound waves that propagated in the hot plasma of tightly coupled photons and
baryons in the early Universe. As the Universe expands and temperature drops
below 3000 K, photons decouple from the baryons at the so called `epoch of
recombination'. The sound speed drops dramatically and oscillatory pattern
imprinted on the baryon distribution as well as the temperature distribution of
the photons which approximately 13 billions years after the Big Bang revealed as
the acoustic oscillations in the temperature anisotropies of the CMB. The
equivalent but attenuated feature exists in the clustering of matter, as baryons
fall into dark matter potential wells after the recombination. In recent years,
the acoustic peak scale in the LSS has been proposed as a potential `standard
ruler' \citep[e.g.][]{Blake03,Glazebrook07,McDonald07} for constraining the Dark
Energy equation of state ($w=p/\rho c^2$) and its evolution.

For the BAO approach to the study of Dark Energy to yield a competitive result, 
a large survey of several million galaxies is generally required
(\citealt{Blake03,Seo03,Parkinson07,Angulo08}). A galaxy spectroscopic redshift
survey would require a substantial amount of time and resources. An alternative
route which will enable a quicker and larger area covered is through the use
photometric redshift (photo--z hereafter) at the expense of the ability to probe
the radial component directly. The photo--z errors are usually worse than
spectroscopic redshift errors, but this can be compensated by a larger survey and deeper
imaging.

The potential of the distribution of Luminous Red Galaxies (LRGs) as a powerful 
cosmological probe has long been appreciated (\citealt{Gladders00,Eisenstein01}). 
Their intrinsically high luminosities provide us with at least two advantages, one
being the ability to observe such a population out to a greater distance whilst
the other is the possibility of detecting the small overdensity of the BAO in
matter distribution at $\approx 100 \hmpc$ owing to their high linear
bias\footnote{This is the well known luminosity dependant bias as shown
observationally by e.g. \cite{Norberg02,Zehavi05b} and is expected in
hierarchical clustering cold dark matter universe \citep{Benson01}.}. In
addition, their typically uniform Spectral Energy Distributions (SEDs) allow a
homogeneous sample to be selected over the volume of interest. Moreover, the
strong 4000 \AA\ break in their SEDs make them an ideal candidate for the
photometric redshift route or even a colour-magnitude cut as demonstrated by the
success of the target selection algorithm of three LRG spectroscopic follow-ups
using SDSS imaging. In fact, the first clear detection of the BAO in the galaxy
distribution came from the analysis of LRG clustering at low redshift
\citep{Eisenstein05}.

Here we shall present new measurements of the angular correlation
functions determined from colour selected LRG samples. We shall show
that this route provides redshift distribution, $n(z)$, widths that 
are close to the current photo--z accuracy, with none of the associated
systematic problems. Indeed, one of our aims is to assess the efficiency
of this route to BAO measurement compared to a full 3--D redshift
correlation function. This possibility arises because the $n(z)$ width
that we obtain is comparable to the $\approx 100 \hmpc$ scale of the
expected acoustic peak.

A similar clustering analysis measuring $w(\theta)$ of LRGs with photo--z's has
been carried out by \cite{Blake08}. Equipped with a higher-redshift LRG
selection algorithm whose effectiveness has been tested with the new LRG
spectroscopic redshift survey, the VST-AA$\Omega$ $ATLAS$ pilot run \citep{Ross08},
our approach is an improvement over \cite{Blake08} as it probes an almost four
times larger cosmic volume and we extend the analysis to large scales to search
for the BAO peak. 

The layout of this paper is as follows. An overview of the galaxy samples used
here is given in \S \ref{sec:data}. \S \ref{sec:method} describes the techniques
for estimating the angular correlation functions and their statistical
uncertainties. We then present the correlation results in \S \ref{sec:result}. 
In \S \ref{evol}, the clustering evolution of these LRGs are discussed. We then 
investigate a possibility of the acoustic peak detection in the $w(\theta)$ 
from the combined sample in \S \ref{sec:bao}. Finally, the summary and conclusions
of our study are presented in \S \ref{sec:conclusion}. 

\begin{table}
	\centering
%	\begin{minipage}{90mm}
	\caption{Summary of the properties of LRG samples used in this study.}
	\begin{tabular}{lcrrc}
        \hline
        \hline
Sample	 & $\zbar$  &Number   &	Density        & Magnitude (AB) \\
         &          &         &$(\rm deg^{-2})$&                \\
        \hline
SDSS	 &    0.35  & 106\,699 & $\approx$13    & $17.5 \le r_{\rm petro} < 19.5 $\\
2SLAQ 	 &    0.55  & 655\,775 & $\approx$85    & $17.5 < i_{\rm deV} < 19.8$\\
AA$\Omega$&   0.68  & 800\,346 & $\approx$105   & $19.8 < i_{\rm deV} \le 20.5$ \\
        \hline
	\hline
	\end{tabular}
%	\end{minipage}
 \label{tab:summary}
\end{table}

\vspace{-4mm}
\section{Data}   
\label{sec:data}
\begin{figure}
	\hspace{-0.3cm}
	\centering
\includegraphics[scale=0.45]{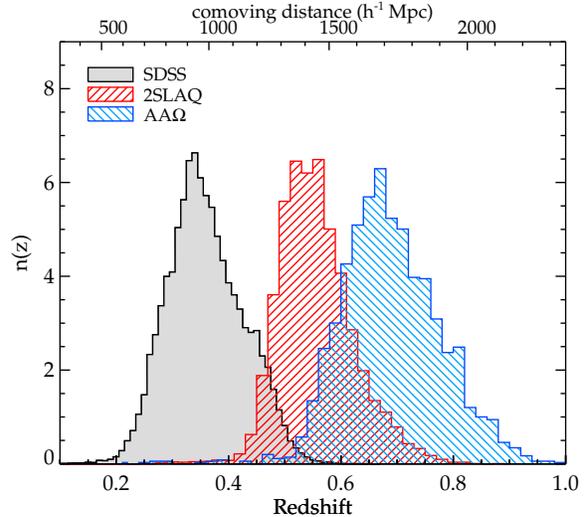}
	\caption{Normalised redshift distributions, $n(z)$, of the three LRG
	spectroscopic surveys used as the basis for selection criteria in this study.}
  \label{nz}
  
\end{figure}

The galaxy samples used in this study were selected photometrically from SDSS
DR5 \citep{Adelman07} imaging data based on three LRG spectroscopic redshift
surveys with $\zbar \approx 0.35, 0.55~\rm{and} ~0.7$
\citep{Eisenstein01,Cannon06,Ross08}. In summary, these surveys utilised a crude
but effective determination of photometric redshift as the strong 4000 \AA\
feature of a typical LRG spectral energy distribution (SED) moves through SDSS
$g,r,i,~\rm{and}~z$ bandpasses \citep{Fukugita96,Smith02}. In each survey, 
a two-colour system (either $g-r$ versus $r-i$ or $r-i$ versus $i-z$) suitable for
the desired redshift range was used in conjunction with r or i-band magnitude to
select luminous intrinsically red galaxies. The method employed by these surveys
has been proven to be highly effective in selecting LRGs in the target redshift
range. The full selection criteria will not be repeated here but a summary of
the algorithms and any additional criteria will be highlighted below (see
\citealt{Eisenstein01,Cannon06,Ross08} for further details). Redshift
distributions, $n(z)$, of the LRGs from the spectroscopic surveys utilised in
this work are shown in Fig. \ref{nz}. The LRG samples corresponding to the
above $n(z)$ have been carefully selected to match our selection criteria hence
these $n(z)$ will be assumed in determining the 3--D correlation functions,
$\xi(r)$, from their projected counterparts, $w(\theta)$, via the Limber 
(\citeyear{Limber53}) equation.

All magnitudes and colours are given in SDSS $AB$ system and are corrected for
extinction using the Galactic dust map of \cite*{Schlegel98}. In this analysis,
we only used the galaxy samples in the most contiguous part of the survey, i.e.
the northern Galactic cap (NGC). All colours described below refer to the
differences in `model' magnitudes \citep[see][for a review on model
magnitudes]{Lupton01} unless otherwise stated.

Hereafter we shall refer to the photometrically selected sample (not to be
confused with the spectroscopic sample from which they are associated) at average
redshift of 0.35, 0.55 and 0.7 as the `SDSS LRG', `2SLAQ LRG' and `AA$\Omega$
LRG', respectively.
 
\subsection{SDSS LRG}
\label{sec:sdss}
\begin{figure}
	\hspace{-0.5cm}
	\centering
        \includegraphics[scale=0.55]{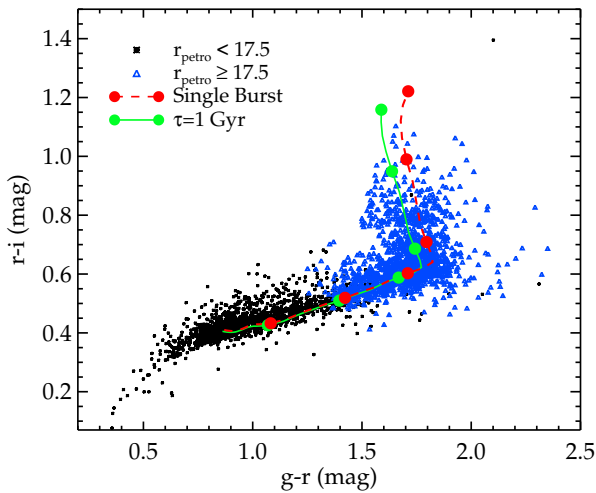}
	\caption[Colour-Colour plot of SDSS LRG]{The colour-colour plot of SDSS LRG
\textit{cut I} and \textit{II} showing their positions on the $gri$ colour plane
compared to the predicted colour-colour locus (observer frame) of typical
early-type galaxies as a function of redshift (see text for more details). Each
solid circle denotes the redshift evolution of the colour-colour tracks at the
interval of 0.1 beginning with $z=0.1$ (bottom left).}
	\label{sdss_track}
\end{figure}

The sample used here is similar to the target sample of the recently completed 
SDSS-LRG spectroscopic survey which contains $\approx100\,000$ spectra and 
cover over $1 ~h^{-3}\,\rm{Gpc}^{3}$. These objects are classified as
LRGs on the basis of their colours and magnitudes following 
\citeauthor{Eisenstein01} (2001, E01 hereafter). The sample is approximately
volume-limited up to $z \approx 0.38$ and spans out to $z \approx 0.5$. The selection
is done using $(g-r)$ and $(r-i)$ colours coupled with r-band \cite{Petrosian76}
magnitude system. The algorithm is designed to extract LRGs in
two different (but slightly overlapped) regions of the $gri$ colour space and
hence using two selection criteria (\textit{Cut I} and \textit{Cut II} in E01)
as naturally suggested by the locus of early-type galaxy on this colour plane
(see Fig. \ref{sdss_track}). The tracks shown in Fig. \ref{sdss_track} were
constructed using a spectral evolution model of stellar populations
\citep{Bruzual03} with output spectra mimicking a typical SED of the LRGs. The
stellar populations were formed at $z \approx 10$  and then evolve with two
different scenarios, namely a) passive evolution of an instantaneous star
formation (single burst), and b) exponentially decayed star formation rate (SFR)
with \textit{e-folding} time of 1 Gyr. Solar metallicity and \cite{Salpeter55}
Initial Mass Function (IMF) were assumed in both evolutionary models. 
  
\begin{figure}
	\hspace{-0.5cm}
	\centering
	\includegraphics[scale=0.5]{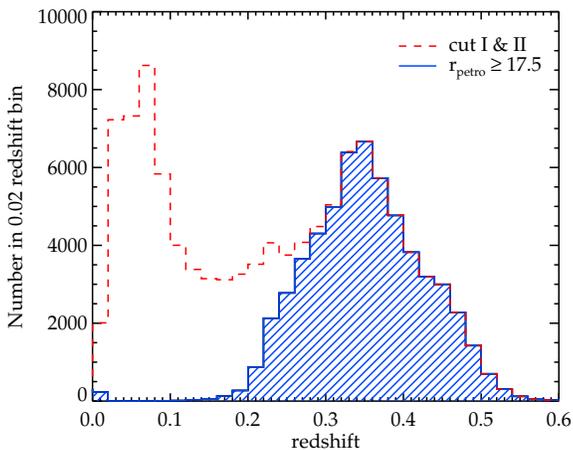}
	\caption{The number of objects as a function of redshift from SDSS LRG
spectroscopic redshift survey also shown is the subset of \textit{Cut I} and
\textit{II} with additional magnitude cut, $\rm{r}_{\rm{petro}} \geqslant 17.5
$, applied.}
	\label{sdss_hist}
	
\end{figure}

We used the same colour-magnitude selection as that described by E01 but with
additional restriction on the $r$-band apparent magnitudes of the objects, i.e.
$r_{\rm{petro}} \geqslant 17.5 $. This is due mainly to two reasons,  a) to
separate out the objects with $z < 0.2 $ because \textit{Cut I} is too
permissive and allows under-luminous objects to enter the sample below redshift
0.2 as also emphasised by E01, and b) to tighten the redshift distribution of
our sample while maintaining the number of objects and its average redshift (see
Fig. \ref{sdss_hist}).

The selection criteria mentioned above also has another star-galaxy separation
algorithm apart from the pipeline PHOTO classification \citep{Lupton01}. This
was done by setting a lower limit on the differences in $r$-band point-spread
function (PSF) magnitudes and model magnitudes as most galaxies populate the
upper part of $r_{\rm PSF}-r_{\rm model}$ space compare to the foreground star
of similar apparent magnitude. The algorithm has been proven to be quite
effective (less than 1 per\,cent stellar contamination) for this range of
redshift and magnitude although \textit{Cut II} needs a more restrictive
threshold, $r_{\rm PSF}-r_{\rm model} > 0.5$ as compared to 0.3 for \textit{Cut
I}. 

In practice, the LRG sample described here can be extracted from the SDSS DR5
imaging database using the SQL query by setting the flag PRIMTARGET to
GALAXY\_RED. This yields a catalogue of approximately $200\,000$ objects which
after applying the additional magnitude cut mentioned above, becomes 106\,699 objects
and results in the sky surface density of about 13 objects per square degree.

\subsection{2SLAQ LRG}
\label{sec:2slaq}
The 2dF-SDSS LRG and Quasar Survey (2SLAQ) is the spectroscopic follow-up of
intermediate to high redshift ($z > 0.4$) LRGs from photometric data of SDSS DR4
\citep{Adelman06} using the two-degree Field (2dF) instrument on the
Anglo-Australian Telescope (AAT). This survey is now completed and contains
approximately $13\,000$ \textit{bona fide} LRGs with over 90 per\,cent at 
$0.45 < z < 0.8$ in two narrow equatorial strips covering 180 square degrees. 
The primary sample of the survey (\textit{Sample 8}, \citealt{Cannon06};C06
hereafter) was selected using $(g-r)$ versus $(r-i)$ colours and `de
Vaucouleurs' $i$-band magnitude $( 17.5 < i_{\rm deV} < 19.8)$. The colour
selection of \textit{Sample 8} is similar to that of \textit{Cut II} which
utilises the upturn of the early-type galaxy locus in $gri$ colour plane and
hence is immune against the confusion with the late-type galaxy locus at higher
redshift (see Fig. 2 in E01) but the scattering up in colour of interlopers
from lower redshift and contamination of M-stars can also affect the accuracy of
the selection. The latter could be prevented by using a similar method for
star-galaxy separation as described in the last section but in this case we used the
$i$-band magnitude rather than the $r$-band. Following C06, two criteria were
used, 

\begin{equation}
i_{\rm psf} - i_{\rm model} > 0.2(21-i_{\rm deV})
\end{equation}
and
\begin{equation}
{\rm radius}_{\rm{deV}(i)} > 0.2, 
\end{equation}
where ${\rm radius}_{\rm{deV}(i)}$ is de Vaucouleurs radius fit of the $i$-band
photometry. As reported by C06, approximately 5 per\,cent of the cool dwarf
M-stars is still present in their sample and we shall assume this value when
correcting for the dilution of the correlation signal due to the uncorrelated
nature of foreground stars and the LRGs. In this work, we only use \textit{Sample
8} as this provides us with a narrower $n(z)$ and higher average redshift than
the whole 2SLAQ sample. 

A sample of 655\,775 photometrically selected LRG candidates ($\approx$ 5 per\,cent
stellar contamination) is returned by the SDSS DR5 `Best Imaging' database when
the \textit{Sample 8} selection criteria is used in the SQL query from table
GALAXY. Objects with BRIGHT or SATURATED or BLENDED but not DEBLENDED flags are
not included in our sample.

\subsection{AA$\Omega$ LRG}
\label{sec:aaomega}
The AA$\Omega$-AAT LRG Pilot observing run was carried out in March 2006 by
\citeauthor{Ross08} (\citeyear{Ross08}, and references therein) as a `Proof of
Concept' for a large spectroscopic redshift survey, VST-AA$\Omega$
\textit{ATLAS}, using the new AAOmega instrument on the AAT.
The survey was designed to target photometrically selected LRGs out to $z \approx
1.0$ with the average redshift of 0.7. The target sample was observed in three
2-degree fields including the COSMOS field \citep{Scoville07}, the COMBO-17 S11
field \citep{Wolf01}, and 2SLAQ d05 field \citep{Cannon06}. 

We follow the survey main selection criteria, $19.8 < i_{\rm deV} \le 20.5$
together with the $riz$ colour cuts as described by \cite{Ross08}. In summary,
the cut utilises the upturn of the early-type galaxy colour-colour locus similar
to that used by 2SLAQ and SDSS LRG surveys. The turning point of the track on
the $riz$ colour plane occurs at $z=0.6-0.7$ as the 4000 \AA\ feature
moves from the SDSS $r$ to $i$ band whilst this happens at $z \approx 0.4$ in
the $gri$ case. The selection technique has been proven to work reasonably well
by the observed redshift distribution. This is further confirmed by the ongoing 
AAT--AA$\Omega$ LRG project, the down-sized version of the VST--AA$\Omega$ 
\textit{ATLAS} survey, designed to observed several thousands of LRG redshifts for 
photo--z calibration and a clustering evolution study. The $n(z)$ (Fig. \ref{nz}) 
used in inferring the 3--D clustering information also includes $\approx$ 2000 AA$\Omega$
LRG redshifts taken during the run in June 2008 (Sawangwit et al. 2011, in prep.).  

As emphasised by \cite{Ross08}, the stellar contamination in the sample can be
readily reduced to $\approx16$ per\,cent by imposing star-galaxy separation
in the $z$-band without any significant loss of genuine galaxies. Although
the level of contamination could be further reduced by using near-infrared
photometry, we do not attempt it here as there is no infrared survey that covers
the entire SDSS DR5 NGC sky with similar depth. Therefore we
shall use the quoted contamination fraction when correcting the measured
$w(\theta)$ for the same reason mentioned in \S \ref{sec:2slaq}. Since no
expression for star-galaxy separation is given in \cite{Ross08}, here such
a procedure is performed using an equation defining the dashed line in their
Fig. 3,

\begin{equation}
z_{\rm psf}-z_{\rm model} > 0.53+0.53(19.0-z_{\rm{model}})
\end{equation}

Applying the above selection rules on the `Best Imaging' data of the SDSS DR5
yields a photometric sample of 800\,346 high-redshift LRG candidates with the sky
surface density of approximately 110 objects per square degree. As with the
2SLAQ LRG sample, objects with BRIGHT or SATURATED or BLENDED but not DEBLENDED
flags are discarded from our sample.

\vspace{-4mm}
\section{Estimating \lowercase{$w(\theta)$} and its error} 
\label{sec:method}
\subsection{Optimal estimator and techniques}
\label{sec:optimal}
The two-point correlation function, $\xi(r)$, measures the excess probability of
finding a pair of objects separated by distance $r$ relative to that expected
from a randomly distributed process. The joint probability of finding two
objects of interest (in this case the LRGs) in the volume elements $\delta
V_{1}$ and $\delta V_{2}$ separated by a distance $r$ is given by
\begin{equation}
	\delta P(r)=n^{2} \left[1+\xi(r)\right] \delta V_{1} \delta V_{2}
	\label{xir_defined}
\end{equation}
where $n$ is the number space density of the sample. In practice, redshift of
individual object is required to estimate the separation between a given pair.
However if such redshift information is not available as in this study, the sky
projected version, $w(\theta)$, can be used to analyse the clustering property
of the sample instead. The 2D equivalent of Eq. \ref{xir_defined} is   
\begin{equation}
	\delta P(\theta)=\aleph^{2} \left[1+w(\theta)\right] \delta \Omega_{1} 
	\delta \Omega_{2}
	\label{angular_defined}
\end{equation}
where $\aleph$ is the surface density of the objects and $\delta P(\theta)$ is
now the joint probability of finding two objects in solid angle $\delta
\Omega_{1}$ and $\delta \Omega_{2}$ separated by angle $\theta$.

Two possible routes for estimating $w(\theta)$ are the pixelisation of galaxy
number overdensity, $\delta_g=\delta_n/\bar{n}$ and pair counting. The
pixelisation approach usually requires less computation time but its smallest
scale probed is limited by the pixel size. We choose to follow the latter. To
calculate $w(\theta)$ using the pair counting method, one usually generates a random
catalogue  whose angular selection function is described by the survey. The
number of random points are generally required to be 10 times the number of
objects or more. This is necessary to reduce the shot noise. Our random
catalogue for each sample has $\approx$ 20 times the number of LRGs in SDSS and
10 times for 2SLAQ and AA$\Omega$-pilot (see next section for details on how
this was achieved).

We compute $w(\theta)$ using the minimum variance estimator of \cite{Landy93}.
It is also an unbiased estimator \citep{Martinez02} for the 2PCF as it can be
reduced to the exact theoretical definition of 2PCF, i.e  a variance of density
fluctuation in Gaussian field, $\xi(r)=\langle \delta(\bf{x})
\delta(\bf{x}+\bf{r})\rangle$. The form of this estimator is
\begin{equation}
w_{\rm LS}(\theta)=1+\left(\frac{N_{rd}}{N}\right)^{2}\frac{DD(\theta)}{RR(\theta)}
-2\left(\frac{N_{rd}}{N}\right) \frac{DR(\theta)}{RR(\theta)}
\end{equation}
where $DD(\theta)$ is the number of LRG-LRG pairs with angular separation within
the angular bin centres at $\theta$. $DR(\theta)$ and  $RR(\theta)$ are the
numbers of LRG-random and random-random pairs, respectively. The $N_{rd}/N$ ratio is
required for normalisation. $N_{rd}$ is the total number of random points and
$N$ is the total number of LRGs. We use a logarithmic bin width of $\Delta
\log(\theta/\rm{arcmin})=0.176$ for $\theta=0.1'$ to $50'$ and a linear bin width
of 20$'$ at scales larger than $50'$. 

The uncertainty in the number density of the sample could lead to a bias in the
estimation of $w(\theta)$ when using Landy-Szalay estimator especially at large
scales where the amplitude is small and hence we also utilise the Hamilton
(\citeyear{Hamilton93}) estimator, given by  
\begin{equation}
  w_{\rm{HM}}(\theta)=\frac{DD(\theta)\cdot RR(\theta)}{DR(\theta)^{2}}-1
\label{hameq}
\end{equation}
which requires no normalisation. We used the Hamilton estimator to cross-check
our $w_{\rm{LS}}$ for each sample and found the difference given by the two
estimators to be negligible in all three samples.   

For the purpose of determining statistical uncertainty in our measurement, three
methods of estimating the errors are considered. The first method is the simple
Poisson error given by 

\begin{equation}
\sigma_\mathrm{Poi}(\theta)=\frac{1+w(\theta)}{\sqrt{DD(\theta)}}
\label{equ:xierr_Poisson}
\end{equation} 
For the second method, {\it field-to-field} error, we split the sample into 24
subfields of approximately equal size. These subfields are large enough for
estimating the correlation function up to the scale of interest. This is simply
a standard deviation of the measurement in each subfield from the best estimate
and is calculated using
\begin{equation}
\sigma_\mathrm{FtF}^{2}(\theta)=\frac{1}{N-1} \sum_{i=1}^{N}
                           \frac{DR_{i}(\theta)}{{DR(\theta)}}  
			   [w_{i}(\theta) - w(\theta)]^{2}
\label{equ:xierr_FtF}
\end{equation}
where $N$ is the total number of subfields, $w_{i}(\theta)$ is a measurement
from the $i$th subfield and $w(\theta)$ is measured using the whole sample. 
The deviation of the angular correlation function computed in each subfield is
weighted by $DR_{i}(\theta)/DR(\theta)$ to account for their relative sizes.  

The third method is the {\it jackknife} resampling. This is a method of
preference in a number of correlation studies \citep[see e.g.][]{Scranton02,
Zehavi05a,Ross07}. The jackknife errors are computed using the deviation of
$w(\theta)$ measured from the combined 23 subfields out of the 24 subfields. The
subfields are the same as used for the estimation of
\textit{field-to-field} error above. $w(\theta)$ is calculated repeatedly, each time 
leaving out a different subfield and hence results in a total of 24 measurements. 
The jackknife error is then 

\begin{equation}
\sigma_\mathrm{JK}^{2}(\theta)=\sum_{i^{\prime}=1}^{N} 
\frac{DR_{i^{\prime}}(\theta)}{{DR(\theta)}}
[w_{i^{\prime}}(\theta) - w(\theta)]^{2}
\label{equ:xierr_jack}
\end{equation}
where $w_{i^{\prime}}(\theta) $ is now an angular correlation function estimated
using the whole sample except the $i$th subfield and
$DR_{i^{\prime}}(\theta)/{DR(\theta)}$ is approximately $23/24$ with slight
variation depending on the size of resampling field. 

The $w(\theta)$ measured from a restricted area are known to suffer from a negative offset 
called `integral constraint', $ic$, which tends to force the fluctuation 
on the scales of the survey to zero \citep{Groth77}, i.e. 
$w_{\rm est}(\theta)= w(\theta)-ic$. The integral constraint can be estimated 
from the random pair counts drawn from the same angular selection function (\S \ref{random}) 
as the data \citep[see e.g.][]{Roche99};
\begin{equation}
ic=\frac{\Sigma RR(\theta)w_{\rm model}(\theta)}{\Sigma RR(\theta)},
\label{equ:w_IC}
\end{equation}
where we assume our fiducial $\Lambda$CDM model (see \S \ref{sec:large}) for 
$w_{\rm model}$. The $ic$ for the SDSS, 2SLAQ and AA$\Omega$-LRG samples are 
$4\times10^{-4}$, $1.5\times10^{-4}$ and $8\times10^{-5}$, respectively. These are much 
smaller than the $w(\theta)$'s amplitudes in the angular ranges being considered in 
this paper, as expected given the large sky coverage of the SDSS data.  

\begin{figure}
	\centering
	\includegraphics[scale=0.43]{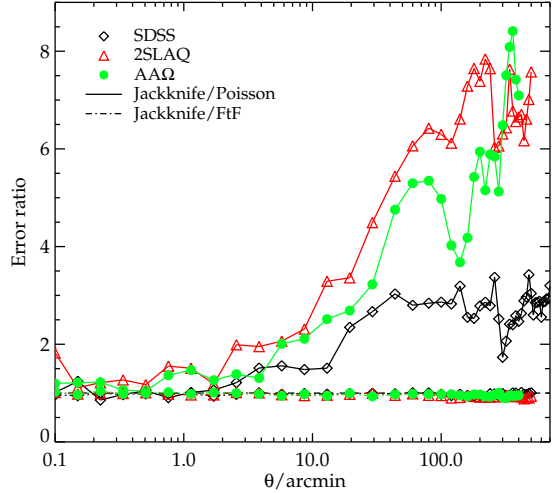}
	\caption{The ratio of jackknife to Poisson and field-to-field errors on 
the measurements of $w(\theta)$. The open diamonds, triangles and solid
circles give the error ratios of $w(\theta)$ estimated from SDSS, 2SLAQ, and
AA$\Omega$ LRG, respectively.}
	\label{err}
\end{figure}

It is well known that the correlation function bins are correlated 
which could affect the confidence limit on the
parameter estimation performed under the assumption that each data point is
independent. Comparison of the estimated error using the field-to-field and
jackknife techniques to the simple Poisson error can give a rough estimate of the
deviation from the independent point assumption. This is plotted in Fig.
\ref{err} which shows that the assumption is valid on small scales where Poisson
error is a fair estimate of the statistical uncertainty. However the same cannot
be said on large scales where the data points are correlated and the independent
point assumption no longer holds. At these scales, such statistical uncertainty
is likely to be dominated by edge-effects and cosmic variance. 

Fig. \ref{err} also shows that the errors estimated using field-to-field and
jackknife method are in good agreement at all angular scales except for 2SLAQ
and AA$\Omega$ samples where the jackknife errors are slightly smaller towards
the large scales but still agree within 10 per\,cent. The errors quoted in later
sections are estimated using the jackknife resampling method. 

\begin{figure*}
	\centering
	\includegraphics[width=170mm]{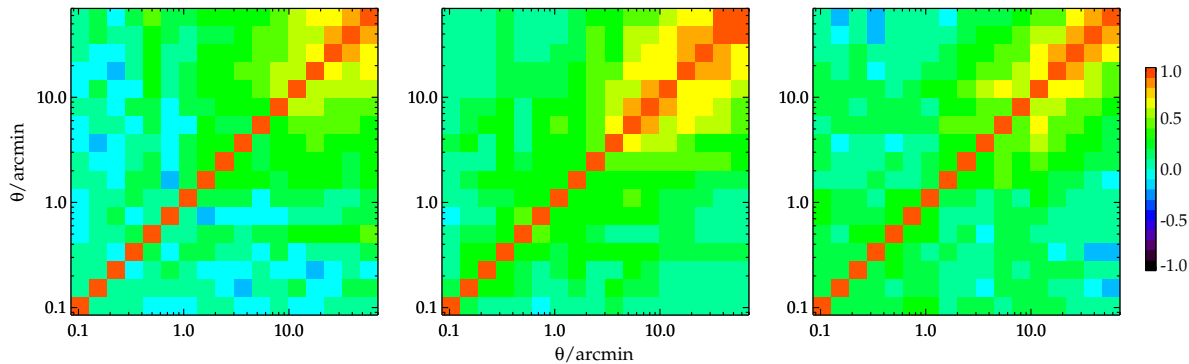} 
	\caption{The correlation coefficients, $\mathbfss{r}_{ij}$, showing the level
of correlation between each angular separation bin for SDSS, 2SLAQ, and	AA$\Omega$
LRG (left to right). Note that for each sample we only show $\mathbfss{r}_{ij}$
up to the angular separation corresponds to $\approx 20 \hmpc$ where later we
shall attempt to fit power-law forms to the measured $w(\theta)$'s.}
	\label{covar}
	\vspace{-4mm}
\end{figure*}

The covariance matrix allows the correlation between each bin to be quantified
and can be used in the fitting procedure to de-correlate the separation bins. We
calculate the covariance matrix from the jackknife resampling using

\begin{equation}
        \mathbfss{C}_{ij} = (N-1) \langle[w(\theta_{i})
- \overline{w(\theta_{i})}]\cdot[w(\theta_{j})-\overline{w(\theta_{j})}]\rangle
\end{equation}
where $\overline{w(\theta_{j})}$ is the mean angular correlation function of all
the jackknife subsamples in the $j$th bin. Note that the difference between
$\overline{w(\theta_{j})}$ and $w(\theta)$ estimated using the whole sample is
negligible. We then proceed to compute the `correlation coefficient', $\mathbfss{r}_{ij}$,
defined by
\begin{equation}
\mathbfss{r}_{ij}=\frac{\mathbfss{C}_{ij}}{\sqrt{\mathbfss{C}_{ii}\cdot
\mathbfss {C}_{jj}}}
\label{eq:corre_coeff}
\end{equation}  
Fig. \ref{covar} shows the correlation coefficients for the three samples
which are strongly correlated at the largest scale considered and less at
small scales confirming the simple correlation test using Poisson errors. 
Note that for the purpose of model fittings in the large-scale sections 
(\S \ref{sec:large}, \ref{sec:HOD} and \ref{sec:bao}) where a more stable 
covariance matrix is required, we increase the number of resampling fields 
to 96 sub-regions with approximately equal area. The size of these sub-regions 
are also big enough for the largest scale being considered in this paper. 
The correlation coefficients constructed from these 96 JK resampling are shown 
in Fig. \ref{fig:corre_large} for the three LRG samples. 

We use the $k$d-trees code \citep{Moore01} to minimise
the computation time required in the pair counting procedure. The angular
correlation function is estimated using the method described above and then
correct for stellar contamination which reduce the amplitude by a factor
$(1-f)^2$, where $f$ is the contamination fraction for each sample given in
\S \ref{sec:data}. 

\begin{figure}
	\centering
	\includegraphics[scale=0.3]{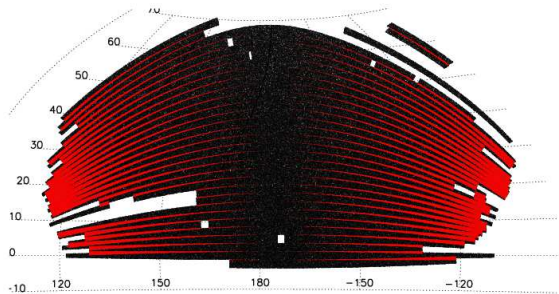}
	\caption{An equal area Aitoff projection of a random catalogue described in 
\S \ref{random}. The red/grey highlighted regions indicate the areas where adjacent
stripes are overlapped. Note that the shading is purely diagrammatic to show the overlap 
regions and is unrelated to galaxy density.}
	\label{overlap}
\end{figure}

\subsection{Constructing random catalogues}
\label{random}
In order to calculate the angular correlation function accurately, a random
catalogue is required. This catalogue consists of randomly distributed
points with the total number at least 10 times that of the data. Each random point 
is assigned a position in Right Ascension (RA) and Declination (DEC). Since our sample 
spans a wide range in DEC (see Fig. \ref{overlap} for the SDSS DR5 sky coverage), care
must be taken to keep the surface number density constant assuming the survey
completeness is constant and uniform throughout. Only the random points that
satisfy the angular selection function of the survey as defined by the mask are
selected. 

The mask is constructed from `BEST' DR5 imaging sky coverage given\footnote{{\tt
http://www.sdss.org/dr5}} in the survey coordinate $(\lambda,\eta)$ and stripe
number. The sky is drift scanned in a strip parallel to $\eta$ and two strips
are required to fill a stripe \citep{York00}. Each stripe is $2.5\deg$ wide and 
their centres are separated by $2.5\deg$. In addition to the `BEST' sky coverage 
mask, we also exclude regions in the quality `holes' and regions defined as 
`BLEEDING', `BRIGHT\_STAR', `TRAIL' and `HOLE' in the `mask' table given by 
the SDSS database. The final mask is applied to both our data and random 
catalogues.

Note that further away from the survey equator ($\rm{RA}_{2000}=185\deg$), the 
adjacent stripes become overlapped which account for almost 20 per\,cent of the 
sky coverage. The `BEST' imaging database only keep the best photometry of the 
objects which have been detected more than once in the overlap regions. 
At the faint magnitude limit of our sample, this could lead to a higher
completeness in the overlap region and introduces bias in the estimated
correlation function. This issue has also been addressed by \cite{Blake07}. They
compared the measurement from the sample which omits the overlap region against
their best estimate and found no significant difference. We follow their
approach by excluding the overlap regions and re-calculating the angular
correlation function of our faintest apparent magnitude sample,
AA$\Omega$-LRG, where the issue is expected to be the most severe. We found
no significant change compared to our best estimate using the whole sample.    

\vspace{-4mm}
\subsection{Inferring 3--D clustering}
\label{sec:limber}
The angular correlation function estimated from the same population with the
same clustering strength will have a different amplitude at a given angular scale
if they are at different depths (redshifts) or have different redshift selection
functions, $\phi(z)$. Therefore in order to accurately compare the clustering
strengths of different samples inferred from $w(\theta)$, one needs to know the
sample $\phi(z)$. Even if the redshifts of individual galaxies are not
available, their 3--D clustering information can be recovered if the sample
redshift distribution, $n(z)$, is known. The equation that relates the spatial
coherence length, $r_{0}$, to the amplitude of $w(\theta)$ is usually referred to
as Limber's equation.

\begin{figure}
\hspace{-0.5cm}
	\centering
	\includegraphics[scale=0.37]{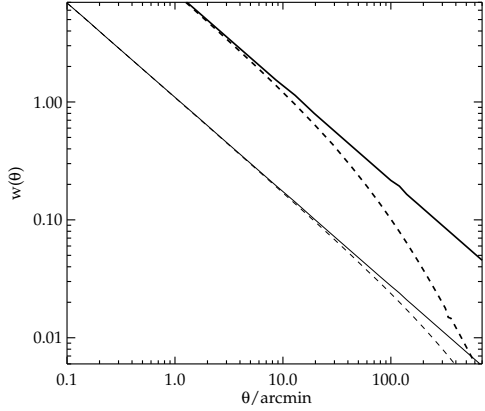}
	\caption{The angular correlation function computed using the full
(dashed-lines) and approximate (solid-lines) Limber equation, derived using a power-law,
$\xi(r)=(r/r_0)^{-\gamma}$ where $r_{0}=10 \hmpc$ and $\gamma=1.8$ with the SDSS
LRG $n(z)$ for the thin lines and much narrower $n(z)$ ($\pm 0.01$ centred at
$z=0.35$) for the thick lines.}
	\label{newlim}
\end{figure}

Recently, the accuracy of Limber's equation has been called into question. This
is due to the assumption made for Limber's approximation that the selection
function, $\phi(z)$, varies much more slowly than $\xi(r)$ in addition to the
flat--sky (small angle) approximation. It was shown by \cite{Simon07} that such
an assumption would lead to $w(\theta)$ being overestimated at large angle where
the breakdown scale becomes smaller for narrower $\phi(z)$ (see Fig.
\ref{newlim}). Here, we shall use the relativistic generalisation of Limber's
equation suggested by \cite{Phillipps78} but without the approximation mentioned
above. Following \cite{Phillipps78} for the comoving case, 

\begin{equation}
w(\theta)=\frac{\int_{0}^{\infty}{dz_{1}f(z_{1})}\int_{0}^{\infty}{dz_{2}f(z_2)
\xi(r)}}{\left[\int_{0}^{\infty}dzf(z)\right]^2}
\label{mylim}
\end{equation}
The source's radial distribution, $f(z)$, is simply given by the galaxy
selection function, $\phi(z)$, as 
\begin{equation}
f(z) \equiv \chi^2(z)\frac{d\chi(z)}{dz}n_c(z)\phi(z)
\end{equation}
where $\chi$ is the radial comoving distance, $n_c(z)$ is the comoving number
density of the sources and $r=r(\theta,z_1,z_2)$ is a comoving separation of the
galaxy pair. We shall assume a spatially flat cosmology (see \S \ref{sec:large}) hence
\begin{equation}
r \equiv \sqrt{\chi^2(z_1)+\chi^2(z_2)-2\chi(z_1)\chi(z_2)\cos \theta}
\end{equation}
Note that Eq. \ref{mylim} can also be used to relate a non-power-law spatial
correlation function to $w(\theta)$ unlike the conventional power-law
approximation of Limber's equation \citep{Phillipps78}.

Fig. \ref{newlim} shows $w(\theta)$ computed using Eq. \ref{mylim} (dashed
lines) compared to the conventional Limber's approximation (solid lines)
for a power-law $\xi(r)$ with clustering length $10 \hmpc$ and $\gamma=1.8$. The
effect of a much narrower redshift distribution (thick lines) is also shown
where the break scale becomes smaller and the power-law slope of $w(\theta)$
asymptotically approaches that of $\xi(r)$, agreeing with the finding of
\cite{Simon07}. We shall use Eq. \ref{mylim} together with the known $n(z)$
to infer the 3--D spatial clustering of the LRGs.  

\vspace{-4mm} 
\section{Results} 
\label{sec:result}
\subsection{Power-law fits} 
\label{sec:result_small}
\begin{figure}
        \hspace{-0.45cm}
	\centering
	\includegraphics[scale=0.53]{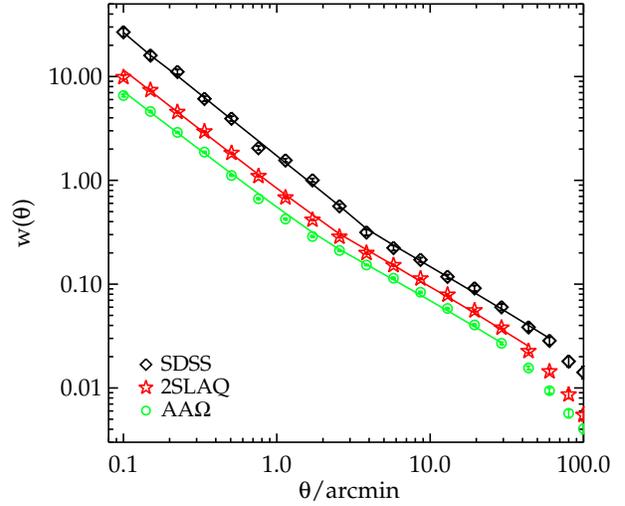}
	\caption{The angular correlation function measured from the three LRG samples.
The solid lines are the projection of best-fit double power-law $\xi(r)$ with
$r_{0}$ and $\gamma$ given in Table \ref{tab:fit} for each sample. The break
scales occur at approximately a few arcminutes depending on the average redshift
of the sample. This corresponds to a comoving separation of $\approx 1 \hmpc$
(see Fig. \ref{fit}).}
	\label{smallall}
\vspace{-4mm}
\end{figure}

\begin{figure*}
	\centering
	\includegraphics[scale=0.55]{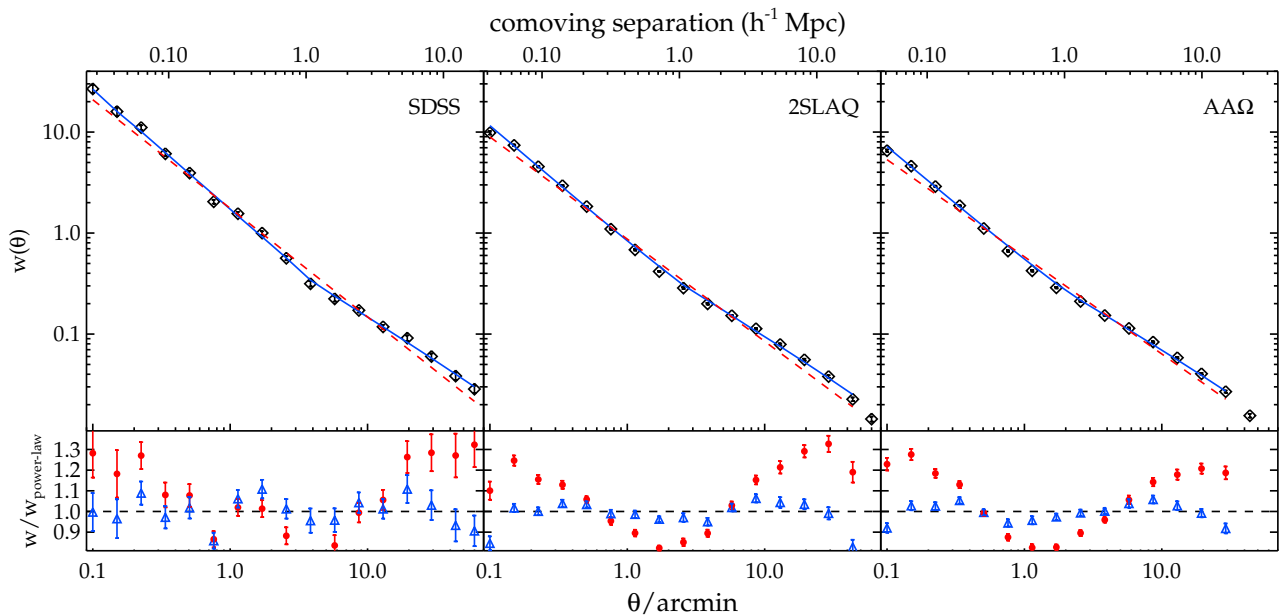}
	\caption{The angular correlation function with the best-fit single 
(red dashed line) and double (blue solid line) power-law for the SDSS, 2SLAQ
and AA$\Omega$ LRGs. Lower panels show the fitting residuals for the single
(circles) and double (triangles) power-law.}
	\label{fit}
\end{figure*}

We first look at the angular correlation function measured from the LRG sample
at scales less than $1\deg$ corresponding to approximately $20 \hmpc$
where previous studies suggested that the spatial 2PCF can be described by a
single power-law of the form $\xi(r)=(r/r_{0})^{-\gamma}$ (typically
$\gamma=1.8$) and a single power-law $w(\theta)$ with slope $1-\gamma$ is
expected (see Fig. \ref{newlim}). However in this study, we find a deviation
from a single power-law with a break in the slope at $\approx 1 \hmpc$ in all three
samples (less significant for the SDSS LRG). The measurement has a
steeper slope at small scales ($< 1 \hmpc$) and is slightly flatter on scales up
to $\approx 20 \hmpc$ where it begins to drop sharply (see Fig. \ref{smallall}
and Fig. \ref{fit}). The inflexion feature at $\approx 1 \hmpc$ has also been reported
in the spatial and semi-projected, $w_{p}(\sigma)$, correlation function by many authors
\citep[e.g.][]{Zehavi05a,Phleps06,Ross07,Blake08} and detections go back as far as 
\cite{Shanks83}. We shall return to discuss these features 
in the halo model framework (\S \ref{sec:HOD}).

\begin{table*}
	\centering

	\caption{Parameters for the power-law fits to the angular correlation function 
derived from three LRG samples. The best-fit parameters given are defined such
that $w(\theta)=(\theta/\theta_{0})^{1-\gamma}$ and $\xi(r)=(r/r_{0})^{-\gamma}$. 
The parameters for the best-fit double power-law are given in two rows where the $\theta<\theta_b$ 
result is given in the top row. Also given are the corresponding $1\sigma$ error
for each parameter.}
    %   \begin{flushleft}
%\begin{minipage}{120mm}
		\begin{tabular}{lcccccccccc}
        \hline
        \hline
Sample	 & $\bar{z}$& $n_{\rm{g}}$   & \multicolumn{4}{c} {Single power-law}   &	 \multicolumn{3}{c} {Double power-law }  \\
         &        & $(h^3\,\Mpc^{-3})$& $\theta_{0}( ' )$ & $\gamma$& $r_{0}(h^{-1}\,\rm{Mpc})$&$\chi^2_{\rm{red}}$&
       $\theta_{0}( ' )$ & $\gamma$& $r_{0}(h^{-1}\,\rm{Mpc})$&$\chi^2_{\rm{red}}$\\
          \hline
SDSS  &	0.35 &$1.1\times10^{-4}$& $1.69\pm0.03$ &	$2.07\pm0.01$	&  $8.70\pm0.09$&16.2&   $1.57\pm0.05$ &	$2.19\pm0.03$	&  $7.35\pm0.08 $&2.2 \\
      &      &       &              &                   &            &     &   $1.05\pm0.09$ & $1.85\pm0.04$& $9.15\pm0.16 $& \\
\hline
2SLAQ  & 0.55 &$3.2\times10^{-4}$&  $0.87\pm0.01$ &	$2.01\pm0.01$	&  $7.50\pm0.04$&57.5  &   $0.83\pm0.01$ &$2.16\pm0.01$	&  $6.32\pm 0.03$&3.9  \\
      &       &      &                &                 &             &    &   $0.60\pm0.03$ & $1.84\pm0.02$&    $ 7.78\pm0.05$& \\   
\hline  
AA$\Omega$ & 0.68 &$2.7\times10^{-4}$& $0.57\pm0.01$ & $1.96\pm0.01$	& $7.56\pm0.03$&42.8 &  $0.56\pm0.01$ &	$2.14\pm0.01$	&  $ 5.96\pm0.03$& 3.4 \\
           &   &   &                   &                &             &    &  $ 0.38\pm0.02$ & $1.81\pm0.02$&    $ 7.84\pm0.04 $& \\           
	\hline
	\hline 
\end{tabular}
\label{tab:fit}
\end{table*}

If we first consider $w(\theta)$ at scales smaller and larger than the break
point separately, each can be approximately described by a power-law with a slope of
$\approx -1.15$ ($\gamma=2.15$), and $\approx -0.83$ ($\gamma=1.83$), respectively. A
more detailed analysis is performed by fitting a set of models to the measured
$w(\theta)$ using a chi-squared minimisation method with the full covariance matrix
constructed from the jackknife resampling (see \S \ref{sec:optimal}).
This allows us to quantify the significance of the
deviation from the single power-law by comparing its \textit{goodness of fit} to
a double power-law. We proceed by calculating

\begin{equation}
\chi^2= \sum_{i,j=1}^{N} \Delta w(\theta_{i})\mathbfss{C}_{ij}^{-1}\Delta
w(\theta_{j})
\label{eqn:chi2}
\end{equation}
where $N$ is the number of angular bins, $\Delta w(\theta_{i})$ is the
difference between the measured angular correlation function and the model for
the $i$th bin, and $\mathbfss{C}_{ij}^{-1}$ is the inverse of covariance
matrix. 

The single power-law fit is of the form
$w(\theta)=(\theta/\theta_{0})^{(1-\gamma)}$. We also recover the spatial
clustering length, $\rm{r}_0$, and its slope through the fitting via Eq.
\ref{mylim}. For a double power-law, the fitting procedure is performed
separately at the scales smaller and larger than $\theta_b$, corresponding to
$\approx 1 \hmpc$ for all three samples (see Fig. \ref{fit}). The largest scale
considered in the fitting for all cases is $\approx 20 \hmpc$ where a steeper
drop-off of $w(\theta)$ is observed.

In Fig. \ref{fit}, the best-fit power-laws for all three samples are
shown. The summary of the best-fit parameters are given in Table
\ref{tab:fit}. Eq. \ref{mylim} and \ref{eqn:chi2} are then used to find the
spatial clustering lengths and slopes that best describe our $w(\theta)$
results. The best-fit clustering slopes from $r_{0}$-$\gamma$ analysis using
Limber's equation are in good agreement with that from $\theta_{0}$-$\gamma$ and
hence we only report the latter in Table \ref{tab:fit}. If we require continuity 
in the double power-law $\xi(r)$ at the break scale, such a scale can be constrained by 
the pair of best-fit $r_{0}$-$\gamma$'s for each sample. From Table \ref{tab:fit}, 
the double power-law break for the SDSS, 2SLAQ and AA$\Omega$ samples are then 
at $2.2,~1.9$ and $1.3\hmpc$, respectively (see \S \ref{sec:small_evol} for further discussion 
of the possible small-scale evolution of $\xi(r)$). By assuming the $1\hmpc$ break instead 
of aforementioned values, the $w(\theta)$ is underestimated by $\approx10$ per\,cent 
for the SDSS case (less for the other two samples) which is only localised to around 
$\theta_b$. The clustering length (single power-law), $r_{0}$, ranges from 7.5 to 
$8.7 \hmpc$, consistent with highly biased luminous galaxies. Single power-law fits to 
the data can be ruled out at high statistical significance. While the double power-law give 
better fits to the data than the single power-law, their $\chi^2_{\rm{red}}$ values indicate 
that such a model is still not a good fit to the data, given the small 
statistical errors. Nevertheless, to first order, the double power-law fits provide 
a good way of quantifying the spatial clustering strength of the samples via the use 
of Limber's equation.

The best-fit slopes at small scales show a slight decrease with increasing redshift, similar 
to that found by \cite{Wake08}. The SDSS LRG sample is more strongly
clustered than the rest as expected. This is simply because the SDSS LRG sample is
intrinsically more luminous than the 2SLAQ and AA$\Omega$ LRG samples and is not an
indication of evolution.  

The galaxy number density (see Table \ref{tab:fit}) are calculated from 
the unnormalised $n(z)$, assuming the redshift distribution from the spectroscopic 
surveys as described in \S \ref{sec:data}. This is galaxy pair-weighted by $n^2(z)$
\cite[see e.g.][]{ARoss09}

\begin{equation}
n_{\rm{g}}= \int {\rm d}z\,\frac{H(z)n(z)}{\Omega_{\rm{obs}}c\chi^2(z)}\times n^2(z) \Big/ \int {\rm d}z\,n^2(z)
\label{eqn:ng}
\end{equation}
where $\Omega_{\rm{obs}}$ is the observed area of the sky, $\chi(z)$ is the comoving distance to 
redshift $z$ and $c$ is the speed of light. The samples' pair-weighted average redshifts determined
in the similar manner as $n_{\rm{g}}$ are consistent with their median redshifts and 
are given in Table \ref{tab:fit}.

\begin{table*}
\centering
	\caption{Properties and the best-fit parameters for double power-law of
$w(\theta)$ measured from the SDSS-density matched samples.}
	
	\begin{tabular}{lccccccc}

	\hline
        \hline
Sample & number & magnitude & $\bar{z}$ & $n_{\rm{g}}$  &\multicolumn{3}{c}
{Double power-law }  \\
       &        &           &     & $(h^3\,\Mpc^{-3})$&$\gamma$ &
        $r_{0}(h^{-1}\,\Mpc)$ & $\chi^2_{\rm{red}}$    \\
          \hline
2SLAQ$^{\ast}$& 182\,841& $17.5< i_{\rm{deV}}<19.32$&    0.53   &  $1.2\times10^{-4}$  &  $2.25\pm0.02$  &$6.33\pm0.04$ & 2.1 \\
            &      &                                                &               &         &  $1.80\pm0.02$  &$8.88\pm0.08$&   \\
\hline
AA$\Omega^{\ast}$& 374\,198& $19.8< i_{\rm{deV}}<20.25$ &  0.67 & $1.1\times10^{-4}$   &	 $2.20\pm0.02$  &$6.25\pm0.03$& 1.7   \\
             &        &                                 &          &                          &  $1.76\pm0.03$  &$9.08\pm0.06$&    \\
                                                                                                                 
	\hline
	\hline                                                                                                          
\end{tabular}

 \label{tab:fit2}
\end{table*}

\begin{figure}
 \hspace{-0.8cm}
  \centering
	\includegraphics[scale=0.53]{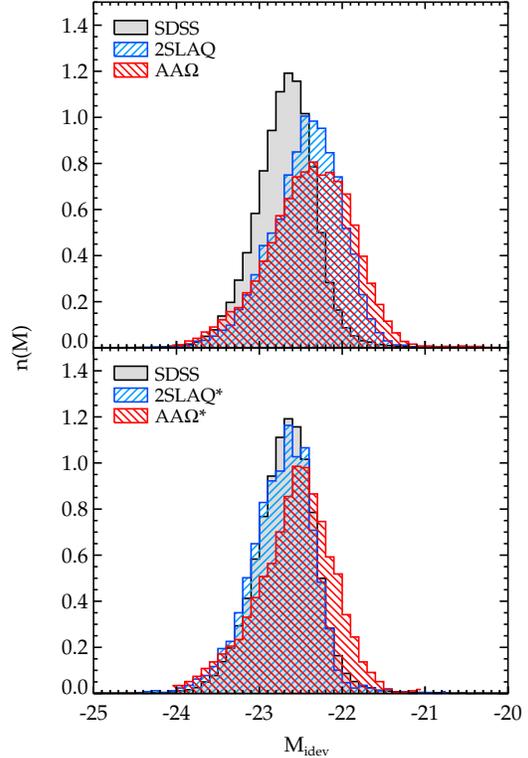}
	\caption{\textit{Top:} The $i$-band absolute magnitude distribution of the
spectroscopic LRG catalogues. All photometry is galactic-extinction corrected
using dust map of \citet*{Schlegel98} and $K+e$ corrected to $z=0$ using the
Early-type galaxy templates from \citet*{Bruzual03}. \textit{Bottom:}
The distribution of the absolute magnitude after applying a faint limit cut to
2SLAQ and AA$\Omega$ LRG in order to match the comoving number density of the
SDSS LRG.}
	\label{fig:abs_mag}
\end{figure}

To this end, we cut back the faint magnitude limit of 2SLAQ and AA$\Omega$ LRG's
to $i_{\rm{deV}}<$ 19.32 and 20.25, respectively. These cuts are imposed in
order to select the samples of galaxies whose comoving number
densities are approximately matched to that of the SDSS LRG. The $K+e$ corrected
$i$-band absolute magnitudes of these samples are presented in Fig. \ref{fig:abs_mag}. 
We see that their absolute magnitudes are also approximately matched. We note that 
we do not attempt to match the LRGs' colour of different samples here. 
This would then allow us to roughly constrain the evolution of LRG clustering up to 
$z \approx 0.68$ (see \S \ref{evol}). A summary of the properties of these samples and the
best-fit parameters are given in Table \ref{tab:fit2}. The measured
$w(\theta)$'s are shown in Fig. \ref{fig:smallreduce}a.

As expected, the amplitudes of the brighter cut 2SLAQ and AA$\Omega$ samples
(denoted by 2SLAQ$^{\ast}$ and AA$\Omega^{\ast}$ hereafter) 
are higher than the original sample. In its raw form, $w(\theta)$ measured from
2SLAQ$^{\ast}$ increases relative to 2SLAQ more than AA$\Omega$ relative to AA$\Omega$*, 
due to the narrower redshift distribution of the 2SLAQ$^{\ast}$ sample. However, if
we perform a double power-law fit to these results, the large-scale, $\ga 1
\hmpc$, clustering lengths are very similar and agree within $\approx1\sigma$
statistical error. To first order these large-scale clustering lengths are also
consistent with that of the SDSS LRG's. We shall investigate the clustering
evolution of these LRG samples further in \S \ref{evol}.

\subsection{Comparison of the clustering form to the standard $\Lambda$CDM model} 
\label{sec:large}

We shall compare our $w(\theta)$ measurements to the
predictions of the standard $\Lambda$CDM model in the linear perturbation theory
of structure growth framework along with the non-linear correction. For the
theoretical models, we first generate matter power spectra, using the `CAMB'
software \citep*{CAMB}. In the case of non-linear correction, the software has
the `HALOFIT' routine \citep{Smith03} implemented. Such matter power spectra,
$P_{\rm{m}}(k,z)$, are then output at the average redshift of each sample. The
matter correlation function, $\xi_{\rm{m}}(r)$, is then obtained by Fourier
transforming these matter power spectra using 

\begin{equation}
\xi_{\rm{m}}(r)=\frac{1}{2\pi^{2}}\int_{0}^{\infty}{P_m(k)k^{2}\frac{\sin{kr}}{kr}}dk
\label{eq:pkxirtransform}
\end{equation}

Under the assumption that galaxies trace dark matter haloes, the galaxy correlation function, 
$\xi_{\rm{g}}(r)$, is related to the underlying dark matter by the bias factor, $b_{\rm{g}}$, via
\begin{equation}
b_{\rm{g}}^2=\frac{\xi_{\rm{g}}(r)}{\xi_{\rm{m}}(r)}
\end{equation}
Therfore the bias factor is expected to be a function of scale unless galaxies
cluster in exactly the same manner as the dark matter does at all scales.
However, at large scales, i.e. the linear regime, the bias
factor is approximately scale--independent over almost a decade of scales 
\citep{Verde02,ARoss08}. 

\begin{figure*}
\hspace{-0.45cm}
\centering
  
\includegraphics[scale=0.53]{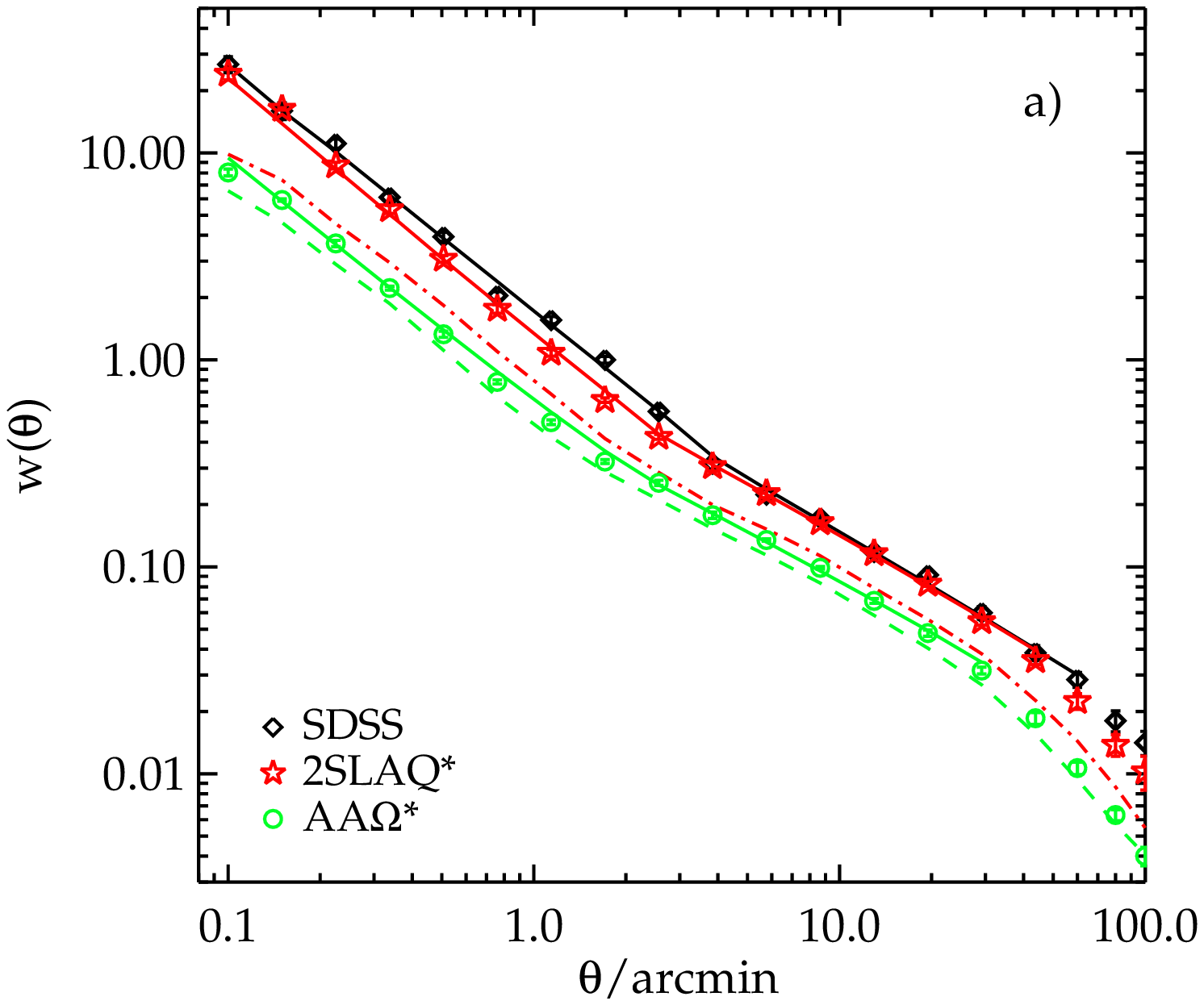}\includegraphics[scale=0.53]{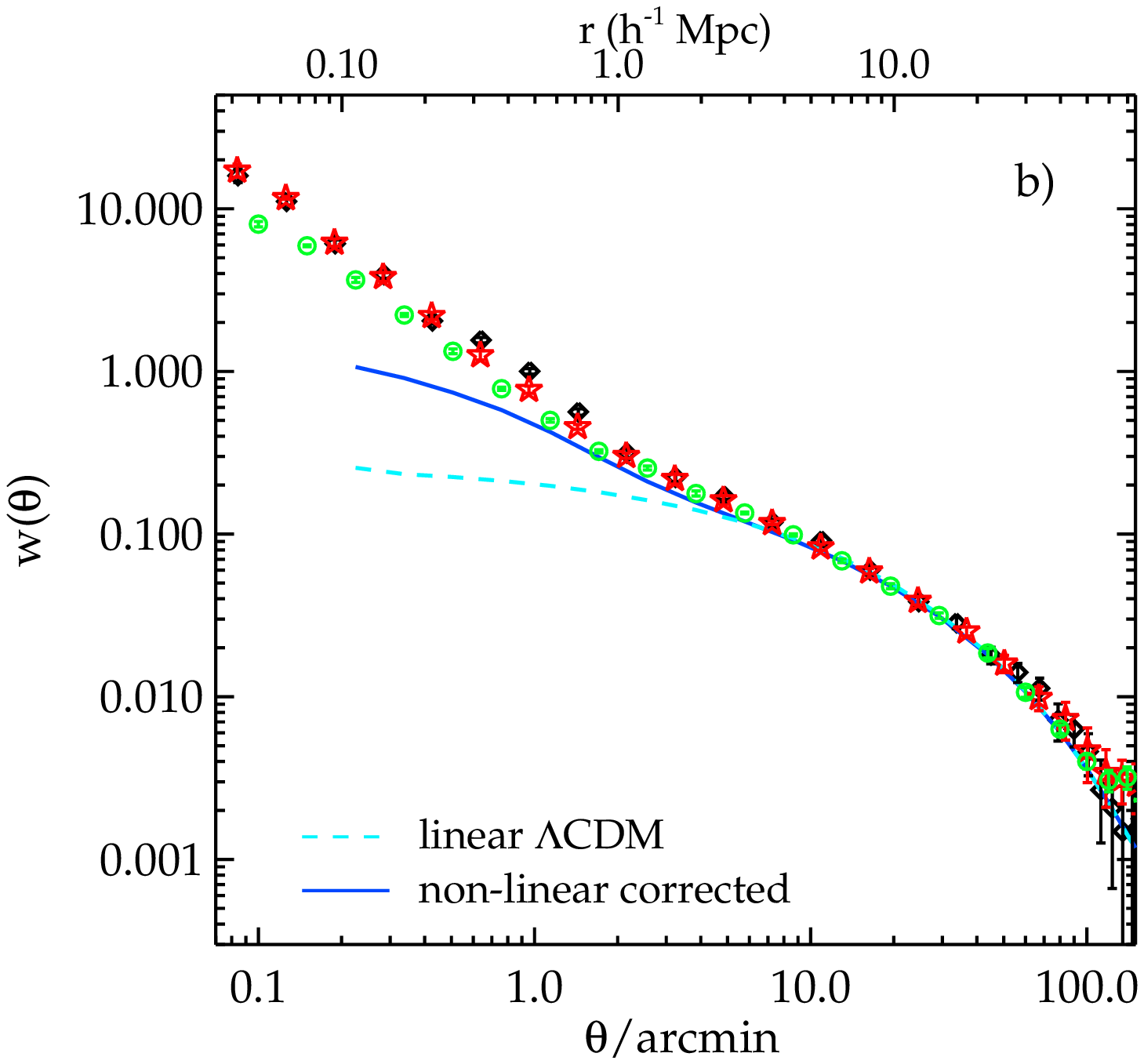}
	\caption{(a): The angular correlation function measured from the
SDSS LRG and the brighter magnitude limit samples drawn from 2SLAQ and
AA$\Omega$ sample (symbols). The solid lines are the projection of the best-fit
double power-law $\xi(r)$ with the parameters shown in Table \ref{tab:fit2}. For
comparison, the dot-dashed and dashed lines are $w(\theta)$ measured from
the whole 2SLAQ and AA$\Omega$ samples, respectively. (b): Same as 
(a) but now scaled to AA$\Omega$ depth and taking into account the
relative amplitude due to the different $n(z)$ widths (see text for more
details).}
	\label{fig:smallreduce}
	\vspace{-4mm}
\end{figure*}

Although we found the clustering lengths and hence the amplitude of $\xi(r)$ 
to be very similar for the SDSS, 2SLAQ$^{\ast}$ and AA$\Omega^{\ast}$ samples,
the evolution in the dark matter clustering means that the linear bias could be
a strong function of redshift as we shall see in the next section where we
investigate the clustering evolution in more detail. The evolution of structures
in linear theory framework is described by the linear growth factor, $D(z)$,
(e.g. \citealt{Peebles84,Carroll92}) such that 

\begin{equation}
\delta(r,z)=D(z)\delta(r,z=0), 
\end{equation}
recall that $\xi(r)=\left\langle \delta(\bf{r_1})\,\delta(\bf{r_2})\right\rangle $, where $r=|\bf{r_1}-\bf{r_2}|$, then
\begin{equation}
\xi_m(r,z)=D^{2}(z)\xi_m(r,0)
\label{eqn:growth}
\end{equation}

The linear growth factor is unity at the present epoch, by definition, and 
decreases as a function of redshift. The $\xi_m(r,z)$ therefore decreases 
as the redshift increases hence given that the number-density/luminosity
matched samples have similar $\xi_g(r)$ amplitudes suggests that the bias 
increases as a function of redshift.

We proceed by projecting the predicted $\xi_{\rm{m}}(r)$ using Eq. \ref{mylim}.
Our fiducial models assume a $\Lambda$CDM Universe with $\Omega_{\Lambda}=0.73,
~\Omega_{\rm{m}}=0.27,~f_{\rm{baryon}}=0.167,~\sigma_{8}=0.8$, $h=0.7$ and 
$n_s=0.95$. The linear bias factor is then estimated by fitting the matter 
$w(\theta)$ to our measurements for the comoving separation of $\approx$
6--60$\hmpc$, using the full covariance matrices. The best-fit linear bias 
($\chi^2_{\rm{red}}$) for SDSS, 2SLAQ$^{\ast}$, AA$\Omega^{\ast}$, 2SLAQ and 
AA$\Omega$ samples are $2.09\pm0.05~(1.2), 2.20\pm0.04~(0.65), 2.33\pm0.03~(0.66), 
1.98\pm0.03~(0.53)$ and $2.07\pm0.02~(1.2)$, respectively. 
The measured biases are consistent with the results from other authors. 
For example, \cite{Tegmark06} analysed $P(k)$ of SDSS LRG and found 
$b(z=0.35)=2.25\pm0.08$ for the best-fit $\sigma_8=0.756\pm0.035$ and for our 
fiducial $\sigma_8$ this becomes $b=2.12\pm0.12$. \cite{Ross07} found 2SLAQ LRG 
$b=1.66\pm0.35$ using redshift-space distortion analysis. \cite{Padmanabhan07} 
, using $C(l)$ of SDSS+2SLAQ photo-z sample, found that $b(z=0.376)=1.94\pm0.06$ 
and $b(z=0.55)=1.8\pm0.04$ (assumed $\sigma_8=0.9$), for our fiducial $\sigma_8$ 
these are $b=2.18\pm0.07$ and $b=2.02\pm0.05$, respectively.

Fig. \ref{fig:smallreduce}b shows the full scaling of of the $w(\theta)$'s, 
accounting for their survey differences. First, the $w(\theta)$ of the SDSS,
and 2SLAQ$^{\ast}$ samples scaled in the angular direction
according to their average redshifts and relative to the AA$\Omega^{\ast}$
sample. The amplitudes are then scaled to obtain a fair comparison for samples
with different redshift distributions. This is done by taking the relative
amplitudes of the projections of a power-law $\xi(r)$ of the same clustering
strength but projected through different $n(z)$ widths. Since the observed
large-scale clustering lengths are very similar, $\approx 9 \hmpc$, the scaled
$w(\theta)$'s in these ranges agree reasonably well. The figure also shows the
best-fit biased non-linear model for the AA$\Omega^{\ast}$ sample.
Our $w(\theta)$ shapes in the ranges $6 \la r \la 60 \hmpc$ can be described
very well by the perturbation theory in the standard flat $\Lambda$CDM
Universe (see the $\chi^2_{\rm{red}}$ for the best-fit bias factor given above).
However, at smaller scales the theory underestimates the clustering amplitude, 
as expected for early-type galaxies. As we shall see in \S \ref{sec:HOD} that 
the reason for this may lie in the details of how the LRGs populate their dark matter 
halo hosts.

\subsection{Halo model fits}
\label{sec:HOD}
\begin{table*}
	\centering

	\caption{Best-fit HOD parameters.}
    %   \begin{flushleft}
%\begin{minipage}{120mm}
		\begin{tabular}{lccccccccc}
        \hline
        \hline
Sample	 & $\bar{z}$ & $M_{\rm min}$ & $M_1$& $\alpha$& $n_{\rm{g}}$   &  $M_{\rm eff}$ & $F_{\rm sat}$ & $b_{\rm lin}$ & $\chi^2_{\rm red}$  \\
       &    &($10^{13}h^{-1}M_{\odot}$)&($10^{13}h^{-1}M_{\odot}$)&  &$(10^{-4}h^3\,\Mpc^{-3})$&($10^{13}h^{-1}M_{\odot}$)& (per\,cent)& &   \\
          \hline
SDSS  &	0.35& $2.5\pm0.2$ & $29.5\pm2.5$& $1.58\pm0.04$&$1.3\pm0.4$&$6.4\pm0.5$& $8.1\pm1.8$ &	$2.08\pm0.05$	& 3.1 \\
2SLAQ* & 0.53& $2.2\pm0.1$ & $27.3\pm2.0$& $1.49\pm0.03$&$1.3\pm0.3$&$4.7\pm0.2$& $7.0\pm0.8$ &	$2.21\pm0.04$	& 7.7 \\
AA$\Omega$*& 0.67& $2.1\pm0.1$ & $23.8\pm2.0$& $1.76\pm0.04$&$1.2\pm0.2$&$4.3\pm0.2$& $5.7\pm0.7$ &$2.36\pm0.04$& 10.1 \\
2SLAQ & 0.55& $1.10\pm0.07$ & $13.6\pm1.1$& $1.42\pm0.02$&$3.2\pm0.5$&$3.4\pm0.2$& $10.0\pm1.1$ &$1.97\pm0.03$	& 14.2 \\
AA$\Omega$& 0.68& $1.02\pm0.03$ & $12.6\pm1.0$& $1.50\pm0.03$&$3.1\pm0.4$&$3.0\pm0.1$& $9.0\pm0.09$ &$2.08\pm0.03$& 13.6 \\
        \hline
	\hline 
\end{tabular}
\label{tab:HODfit}
\end{table*}

We fit a halo model \cite[e.g.][]{Peacock00,Berlind02,Cooray02} to our 
angular correlation function results. 
One of the key ingredients of the halo 
model is the Halo Occupation Distribution (HOD) which tells us how the galaxies 
populate dark matter haloes as a function of halo mass. Recently, 
the model has been used to fit various datasets as a means to physically 
interpret the galaxy correlation function and gain insight into their evolution 
\citep[e.g.][]{White07,Blake08,Wake08,Brown08,ARoss09,Zheng09}. 

Here, we use a three-parameter HOD model \cite[e.g.][]{Seo08,Wake08} which distinguishes 
between the central and satellite galaxies in a halo \citep{Kravtsov04}. The mean number 
of galaxies residing in a halo of mass $M$ is 
\begin{equation}
\left<N(M)\right>=\left<N_{\rm c}(M)\right>\times(1+\left<N_{\rm s}(M)\right>),
\label{eq:Ngal} 
\end{equation}
where the number of central galaxy is either zero or one with the mean given by
\begin{equation}
\left<N_{\rm c}(M)\right>={\rm{exp}}\left(\frac{-M_{\rm{min}}}{M}\right).
\label{eq:Ncen} 
\end{equation} 
We assume that only haloes with a central galaxy are allowed to host satellite galaxies.
In such a halo, the satellite galaxies are distributed following an NFW profile \citep{NFW97} 
around a central galaxy at the centre of the halo. We also assume that their numbers follow 
a Poisson distribution \citep{Kravtsov04} with a mean
\begin{equation}
\left<N_{\rm s}(M)\right>=\left(\frac{M}{M_1}\right)^\alpha
\label{eq:Nsat}
\end{equation}

The NFW profile is parametrised by the concentration parameter $c\equiv r_{\rm vir}/r_{\rm s}$ 
where $r_{\rm vir}$ is the virial radius and $r_{\rm s}$ is the characteristic scale radius. 
We assume \cite{Bullock01} parametrisation of the halo concentration as a function of mass and 
redshift, 
\begin{equation}
c(M,z)\approx\frac{9}{(1+z)}\left(\frac{M}{M_{\ast}}\right)^{-0.13},
\label{eq:concentration}
\end{equation}
where $M_{\ast}$ is the typical collapsing mass and is determined by 
solving Eq. \ref{equation:sig_m_int} with $\sigma(M_{\ast})=\delta_c(0)$.

The galaxy number density predicted by a given HOD is then 
\begin{equation}
n_{\rm{g}}=\int {\rm{d}}M\,n(M)\left<N(M)\right>
\label{eq:ng}
\end{equation}
where $n(M)$ is the halo mass function, here we use the model given 
by \cite{Sheth99}. The effective galaxy linear bias can be determined 
from the HOD;
\begin{equation}
b_{\rm lin}= \frac{1}{n_{\rm{g}}}\int {\rm{d}}M\,n(M)b(M)\left<N(M)\right>,
\label{eq:blin}
\end{equation} 
where $b(M)$ is the halo bias as a function of mass, for which we use the model 
of \cite*{Sheth01} plus the improved parameters of \cite{Tinker05} 
(see \S \ref{sec:halomass}, Eq. \ref{equation:ellipsoidal_bias}).
The average mass of haloes hosting such a galaxy population is then 
\begin{equation}
M_{\rm eff}= \frac{1}{n_{\rm{g}}}\int {\rm{d}}M\,n(M)M\left<N(M)\right>
\label{eq:Meff}
\end{equation}
And the satellite fraction of the galaxy population is given by
 \begin{equation}
F_{\rm sat}= \frac{1}{n_{\rm{g}}}\int {\rm{d}}M\,n(M)\left<N_{\rm c}(M)\right>\left<N_{\rm s}(M)\right>
\label{eq:Fsat}
\end{equation}

The galaxy power spectrum/correlation function can then be modelled 
as having a contribution at small scales that arises from galaxy pairs  
in the same dark matter halo (1-halo term). On the other hand, the galaxy 
pairs in two separate haloes (2-halo term) dominate at larger scales,
\begin{equation}
P(k)= P_{1{\rm h}}+P_{2{\rm h}}
\label{eq:PkHOD}
\end{equation}
The 1-halo term can be distinguished into central-satellite, $P_{\rm cs}(k)$, and
satellite-satellite, $P_{\rm ss}(k)$, contributions \citep[see e.g.][]{Skibba09};
\begin{equation}
P_{\rm cs}(k)= \frac{1}{n^2_{\rm g}}\int {\rm d}M\,n(M) 2\left<N_{\rm c}(M)\right>\left<N_{\rm s}(M)\right>u(k,M),
\label{eq:Pkcs}
\end{equation}
and
\begin{equation}
P_{\rm ss}(k)= \frac{1}{n^2_{\rm g}}\int {\rm d}M\,n(M)\left<N_{\rm c}(M)\right>\left<N_{\rm s}(M)\right>^2u(k,M)^2,
\label{eq:Pkss}
\end{equation}
where $u(k,M)$ is the Fourier transform of the NFW profile and we have simplified 
the number of satellite-satellite pairs $\left<N_{\rm s}(N_{\rm s}-1)\right>$ 
to $\left<N_{\rm s}(M)\right>^2$, i.e. Poisson distribution.

For 2-halo term, we implement the halo exclusion, `$n_{\rm g}'$-matched', and scale-dependent 
halo bias, $b(M,r)$, of \cite{Tinker05};
\begin{eqnarray}
P_{2{\rm h}}(k,r)&=&P_{\rm m}(k) \times \frac{1}{n_{\rm g}'^2} \nonumber \\
     & &\hspace{-14mm}\times\left[\int_0^{M_{\rm lim}(r)} \hspace{-2mm} {\rm d}M\,n(M)b(M,r)\left<N(M)\right>u(k,M)\right]^2,
\label{eq:Pk2h}
\end{eqnarray}
where $P_{\rm m}(k)$ is a non-linear matter power spectrum (see \S \ref{sec:large}), 
$M_{\rm lim}(r)$ is the mass limit at separation $r$ due to halo exclusion and 
$n_{\rm g}'$ is the restricted galaxy number density \citep[Eq. B13 of][]{Tinker05}. 
The scale-dependent halo bias is given by \citep{Tinker05}
\begin{equation}
b^2(M,r)=b^2(M)\frac{\left[1+1.17\xi_{\rm m}(r)\right]^{1.49}}{\left[1+0.69\xi_{\rm m}(r)\right]^{2.09}},
\label{eq:bMr}
\end{equation}
where $\xi_{\rm m}$ is the non-linear correlation function (see \S \ref{sec:large}).

The galaxy correlation function is then the Fourier transform of the power 
spectrum which can be calculated separately for 1- and 2-halo terms. 
For the 2-halo term, we need to correct the galaxy pairs from the restricted galaxy 
density to the entire galaxy population. This is done by
\begin{equation}
1+\xi_{2{\rm h}}(r)=\left(\frac{n_{\rm g}'}{n_{\rm g}}\right)^2\left[1+\xi_{2{\rm h}}'(r)\right],
\label{eq:xi2h}
\end{equation}   
where $\xi_{2{\rm h}}'(r)$ is the Fourier transform of Eq. \ref{eq:Pk2h}.

\begin{figure}
\hspace{-0.45cm}
\centering
\includegraphics[scale=0.5]{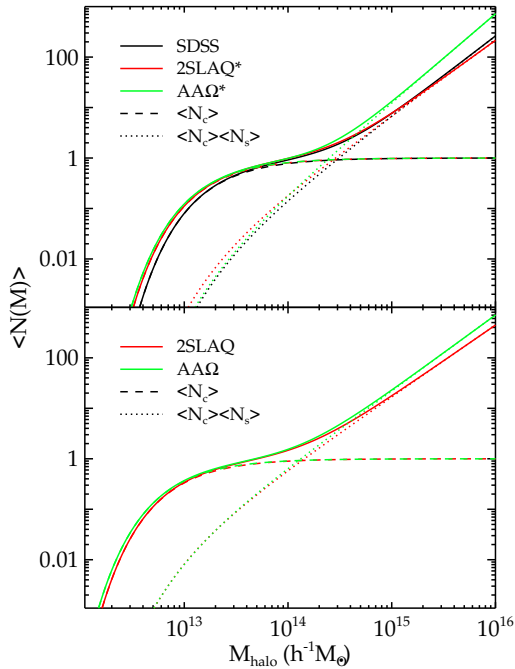}
	\caption{The mean number of LRGs per halo as a function of mass (solid lines) from the best-fit HOD 
	for the SDSS, 2SLAQ*, AA$\Omega$* samples (top) and 2SLAQ, AA$\Omega$ samples (bottom). 
	The central and satellite contributions for each sample are shown as the dashed and dotted lines.}
	\label{fig:HODfit}
	\vspace{-4mm}
\end{figure}

\begin{figure*}
\hspace{-0.45cm}
\centering
  
\includegraphics[scale=0.50]{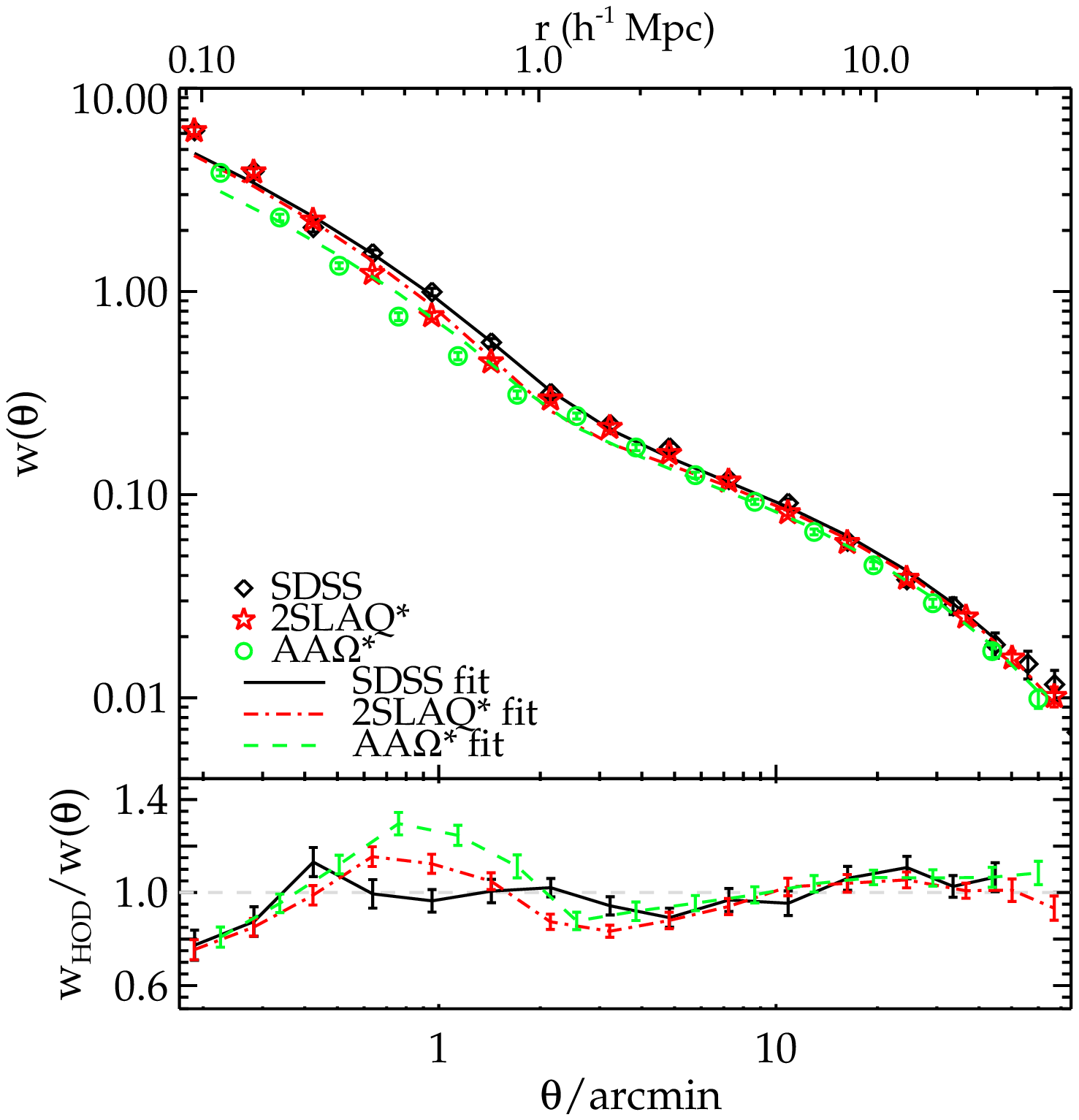}\includegraphics[scale=0.50]{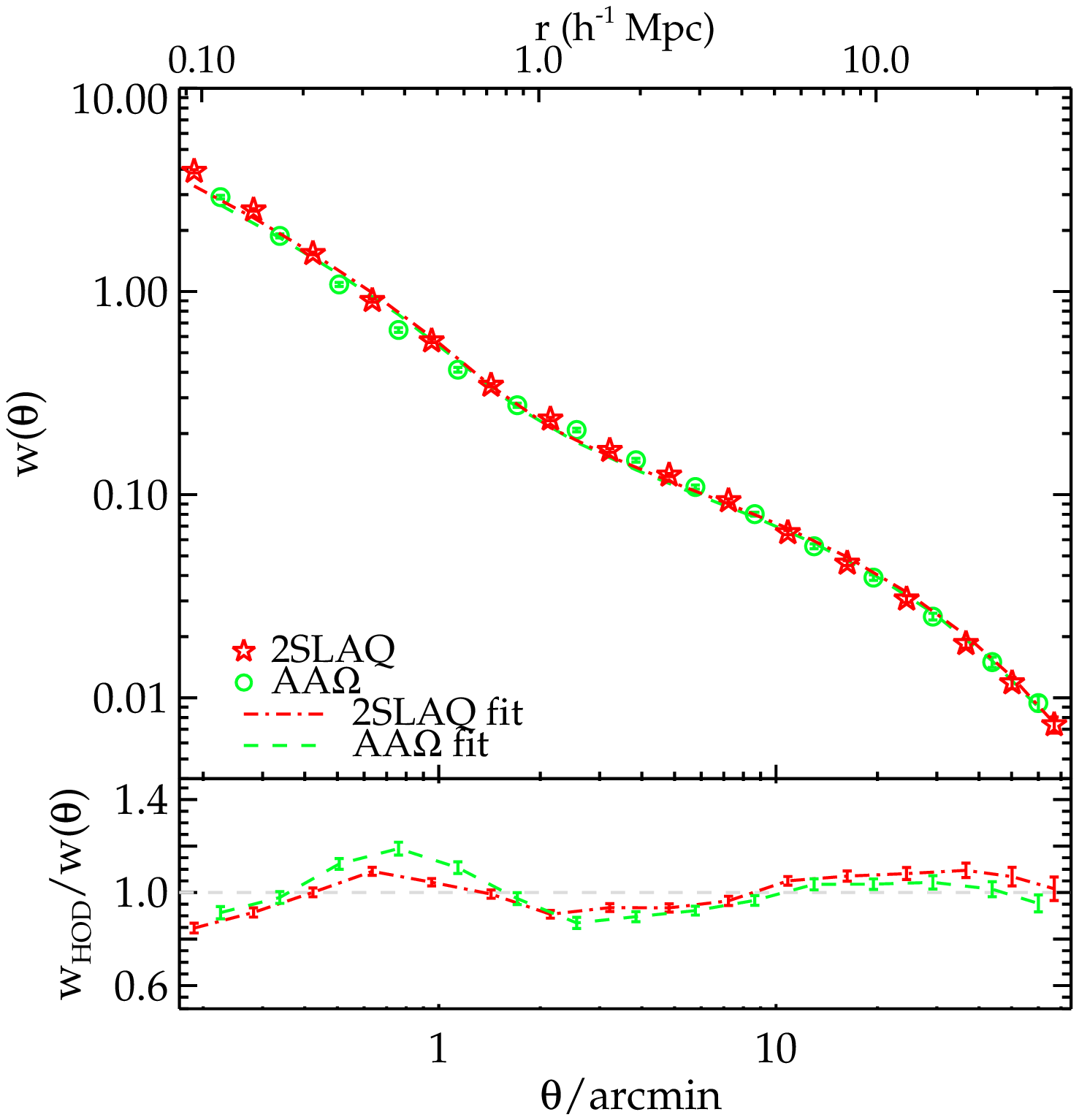}
	\caption{The best-fit HOD models for the SDSS, 2SLAQ*, AA$\Omega$* samples (left)
	and 2SLAQ, AA$\Omega$ samples (right). These are scaled to the AA$\Omega$*/AA$\Omega$ 
	depth similar to that shown in Fig. \ref{fig:smallreduce}b. The bottom panels show the ratios between 
	the best-fit HOD models and the measured correlation functions.}
	\label{fig:HODreduce}
	\vspace{-4mm}
\end{figure*}

We then project the predicted galaxy correlation function to $w(\theta)$ using Eq. \ref{mylim} 
for a range of HOD parameters. The best-fit model for each of our sample is then determined from 
chi-square minimisation using the full covariance matrix. Note that we exclude angular bins 
corresponding to scales smaller than $0.1 \hmpc$ because any uncertainty in the $\xi(r)$ model 
at very small scales, $r \la 0.01 \hmpc$, can have a strong effect on $w(\theta)$ even at 
these scales due to the projection. The best-fit $M_{\rm min}$, $M_{1}$ and $\alpha$ and the 
associated values for $n_{\rm g}$, $M_{\rm eff}$, $F_{\rm sat}$ and $b_{\rm lin}$ are given in 
Table \ref{tab:HODfit}. The $1\sigma$ uncertainties on the best-fit $M_{\rm min}$, $M_{1}$ and 
$\alpha$ are determined from the parameter space where $\Delta \chi^2 \le 1$. For $n_{\rm g}$, 
$M_{\rm eff}$, $F_{\rm sat}$ and $b_{\rm lin}$ which depend on the three main HOD parameters, 
this becomes $\Delta \chi^2 \le 3.53$. Fig. \ref{fig:HODfit} shows the best-fit HOD for each sample, 
the coloured solid lines are the mean number of LRGs per halo with the central and satellite 
contributions shown separately as the dashed and dotted lines, respectively.

As expected, the LRGs populate rather massive dark matter haloes with the masses $\approx10^{13}-10^{14} 
h^{-1} M_{\odot}$. At approximately the same redshift, the more luminous samples, 2SLAQ* and AA$\Omega$*, 
are hosted by more massive haloes than fainter samples. Most of the LRGs, $> 90$ per\,cent, are central galaxies 
in their dark matter haloes, the satellite fraction is only 10 per\,cent or less with the the increasing trend 
towards low redshift. This can be explained in the framework of halo mergers at lower redshift 
(see \S \ref{sec:HODevolution}). The best-fit linear bias factors for all samples are in 
excellent agreement with the values derived in \S \ref{sec:large}. Also the galaxy number density from the best-fit 
halo model is consistent with that derived from Eq. \ref{eqn:ng} (see Tables \ref{tab:fit} and \ref{tab:fit2}).  

Note that, to first order, our best-fit HODs are compatible with the measurements from other authors although 
a direct comparison with samples selected differently may not be simple. For example, our SDSS sample has similar 
space density (although at higher redshift, $z=0.35$ versus 0.3) as the sample studied by \cite{Seo08}. 
Our $M_1/M_{\rm min}$ and satellite fraction are in excellent agreement with their model 11 
(their best-fit $N$-body evolved HOD). But their $\alpha$ is somewhat lower which is caused by the higher 
$\sigma_8=0.9$ value \citep{Wake08} and the lower average redshift. Their $M_1$ and $M_{\rm min}$ are also 
somewhat higher than our best-fit values for the same reason as for $\alpha$. Another example, our best-fit 
$M_{1}$, $M_{\rm min}$, $b_{lin}$ and $F_{\rm sat}$ for 2SLAQ* sample are in good agreement with \cite{Wake08} 
$z=0.55$ 2SLAQ selection, although our values are somewhat higher which may be due to our lower galaxy number 
density, implying that our sample contains rarer and more biased objects.   

The best-fit models for $w(\theta)$ are shown in Fig. \ref{fig:HODreduce}, comparing to the data. Both the models 
and data are scaled to account for the projection effect (see \S \ref{sec:large}) and are 
plotted at the depth of AA$\Omega$*/AA$\Omega$ sample. We immediately see that while the fits at the large scales 
($r \ga 3 \hmpc$) are good, the fits at the small scales and at $r \approx 1-2 \hmpc$ are rather poor especially 
for the higher redshift samples. This is evident in the high best-fit reduced chi-square values in Table 
\ref{tab:HODfit}. Given our small error bars, this may indicate that a more complicated halo model may be needed, 
e.g. five/six parameters HOD, an improved halo-exclusion model \cite[see Fig. 11 of][]{Tinker05}, or different 
halo concentration parametrisation. Another important point to note is that the HOD formalism assumes 
a volume-limited sample, which we do not have here. This means that our observed galaxy number density 
corresponds to a cut-off which evolves with redshift rather than a cut-off in halo mass or LRG luminosity. 
Nevertheless, to first order the HOD fits generally describe the shape and amplitude of our measured $w(\theta)$ 
and we believe that the derived $b_{\rm lin}$ and $M_{\rm eff}$ are reasonably robust despite the statistically 
poor fits. 

\section{Evolution of LRG clustering and dark matter halo masses}
\label{evol}

\subsection{Intermediate scales}
We study the LRGs clustering and dark matter halo mass evolution by employing 
the methods used by \cite{Croom05} and \cite{daAngela08} to analyse their QSO 
samples. We then proceed by considering the small-scale clustering 
evolution in the framework of the halo model. 

\subsubsection{Clustering evolution}
\label{sec:evol_inter}
In this section, we make an attempt to quantify the clustering evolution of the
LRGs via the use of the $w(\theta)$'s measured from the number-density (roughly
luminosity) matched samples as presented in the last section. We shall first 
compare the result at the intermediate scales, $1 \ga r \ga 20 \hmpc$, to the 
simple long--lived model of \citet{Fry96}. The model assumes that
galaxies are formed at a particular time in the past and their clustering
evolution is determined by the influence of gravitational potential where no
galaxies are destroyed/merged or new population created, hence preserving the
comoving number density. In such a model the galaxy linear bias is given by

\begin{equation}
b(z)=1+\frac{b(0)-1}{D(z)}
\label{equa:biasfry}
\end{equation}
and as we saw in \S \ref{sec:large} that $\xi_m(r,z)=D^{2}(z)\xi_m(r,0)$,
the clustering evolution is such that  
\begin{equation}
\xi_g(r,z)=\left[\frac{b(0)+D(z)-1}{b(0)}\right]^2 \xi_g(r,0)
\label{equa:evol_fry}
\end{equation}

We shall also compare the data directly to the linear theory prediction 
for dark matter evolution in the $\Lambda$CDM model, $\xi(r,z) \propto D^2(z)$. 
In addition, we shall
also check the stable clustering and no--evolution (comoving) clustering 
models of \cite{Phillipps78}. The stable model refers to 
clustering that is virialised and therefore stable in proper coordinates. 
For a $\xi(r)$ with $r$ measured in comoving coordinates, the stable model has
evolution $\xi(r)\propto(1+z)^{\gamma-3}$ and the no-evolution model has
$\xi(r)$ independent of redshift. At these intermediate scales, the
clustering is unlikely to be virialised so the stable model is shown
mainly as a reference point. From Eq. \ref{equa:evol_fry}, the no--evolution 
model represents the high bias limit of the long--lived model of \cite{Fry96}. 
The stable and comoving models are similar to the long--lived model in that
they both assume that the comoving galaxy density remains constant with
redshift.

In order to quantify the clustering amplitude of each
sample, we shall use the integrated correlation function 
in a $20\hmpc$ sphere as also utilised by several authors 
\citep[e.g.][]{Croom05,Ross08,daAngela08}. 
The volume normalisation of this quantity is then given by

\begin{equation}
\xi_{20} = \frac{3}{20^3}\int_{0}^{20}\xi(r)r^2dr
\label{equation:xi20}
\end{equation}

The $20 \hmpc$ radius is chosen to ensure a large enough scale for linear theory
to be valid and in our case the power-law with $\gamma\approx1.8$ remains a good
approximation up to $\approx 20 \hmpc$. Furthermore, the non-linearity at small
scales does not significantly affect the clustering measurements, when averaged
over this range of scales.

The integrated correlation function, $\xi_{20}$, approach also provides another
means of measuring the linear bias of the sample. For this, we again assume
scale-independent bias which is a reasonable assumption in the linear regime.
The bias measured in this way is given by

\begin{equation}
b_g(z)=\sqrt{\frac{\xi_{20,g}}{\xi_{20,m}}} 
\end{equation}

The mass integrated correlation functions are again computed assuming
our fiducial cosmological model using the matter power spectra output from CAMB.
The values for $\xi_{20,m}$ used here are 0.153, 0.126 and 0.112 for $z=$ 0.35, 
0.55 and 0.68, respectively.

\begin{figure}
\hspace{-0.7cm}
   \centering
	\includegraphics[scale=0.47]{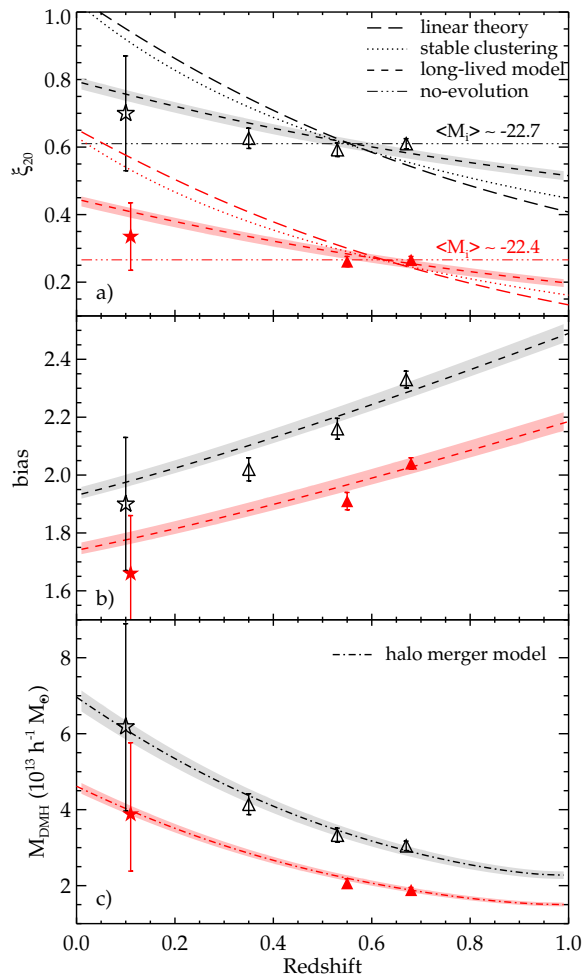}   
   \caption{(a): The LRG $\xi_{20}$ measurements as a function of
redshift and luminosity. The data at $z\approx0.1$ (stars) are taken from the 
correlation functions of early-type galaxise in 2dFGRS \citep{Norberg02}. Open and 
solid symbols correspond to the samples with median absolute
magnitude, $M_{i}-5\log_{10}h= -22.7$ (SDSS/2SLAQ*/AA$\Omega$*) and $-22.4$ 
(2SLAQ/AA$\Omega$). The best fits for various models are also shown (see text for more details). 
The lower luminosity data have been lowered by 0.2 for clarity. (b): The 
LRG linear biases as a function of redshift and luminosity, comparing to the 
best--fit long--lived model. (c): The typical 
mass of dark matter haloes occupied by the LRGs as estimated from the halo bias 
function. The dot-dashed lines are the best-fit evolution model of dark matter 
halo mass via the merger framework \citep{Lacey93}.}   
   \label{fig:evol}
\end{figure}

The $\xi_{20,g}$ is calculated using the best-fit double power-law parameters
for each sample. The results are plotted in Fig. \ref{fig:evol}a
along with the best-fit linear theory evolution (long--dashed
line), stable clustering (dotted line), long-lived (dashed line) and 
no-evolution models (dot-dot-dashed line). The linear bias factors
measured using the $\xi_{20}$ approach are given in Table \ref{tab:evol} and
also presented in Fig. \ref{fig:evol}b. The bias factors determined here are in 
good agreement with the large-scale $\Lambda$CDM (\S \ref{sec:large}) and HOD 
(\S \ref{sec:HOD}) best-fit models.  

To extend the redshift range, we shall compare our results to the clustering of 
early-type galaxies in 2dFGRS studied by \citet{Norberg02} that roughly
match the absolute magnitude of our samples after the $K+e$ correction. These are 
the samples with $-21.0 > M_{b_j}-5\log_{10}h > -22.0$ and 
$-20.5 > M_{b_j}-5\log_{10}h > -21.5$, being compared to the SDSS/2SLAQ*/AA$\Omega$* 
and 2SLAQ/AA$\Omega$ data and denoted N02E1 and N02E2 in Table \ref{tab:evol}, respectively. 
We proceed in a similar fashion to the procedure described above and use 
the author's best-fit power-law to estimate the $\xi_{20,g}$'s and hence the
bias values (see Table \ref{tab:evol}).

Both luminosity bins can be reasonably fitted by the long--lived model. The best-fit 
models for the $M_{i}-5\log_{10}h=-22.7$ and -22.4 samples have $b(0)=1.93\pm0.02$ 
and $1.74\pm0.02$ with $\chi^2=7.34$ (3 d.o.f) and 4.11 (2 d.o.f) respectively, 
i.e. $1.5-1.9\sigma$ deviation. This is interesting given the lack of number
density evolution seen in the LRG luminosity function by \citet{Wake06}. Nevertheless, 
it is intriguing that such a simple model gets so close to fitting data over the wide redshift 
range analysed here.

\begin{table}

	\caption{Summary of the estimated LRG and 2dFGRS early-type galaxy bias factor
and $M_{\rm{DMH}}$ as a function of redshift and luminosity.}
%\begin{minipage}{5cm}
% \hspace{-0.6cm}
%\centering
% \resizebox{!}{2.0cm}
	{\begin{tabular}{lcccc}

	\hline
        \hline
Sample&$\overline{z}$&$\overline{M}_i$&$b$&$M_{\rm{DMH}}$\\%&$b_{\rm{lin}}$\\
      &              &$-5\log_{10}h$&  &$(10^{13}h^{-1}M_{\odot})$\\%&(10-40$h^{-1}\Mpc$)\\          
 \hline
SDSS&0.35&-22.67&$2.02\pm0.04$& $4.1\pm0.3$\\%& $2.10\pm0.04$\\
2SLAQ*&0.53&-22.69&$2.16\pm0.04$&$3.3\pm0.2$\\%& $2.26\pm0.04$\\
AA$\Omega$*&0.67&-22.60&$ 2.33\pm0.03$&$ 3.1\pm0.1$\\%& $2.37\pm0.03$\\
2SLAQ&0.55&-22.40&$ 1.91\pm0.03$&$ 2.1\pm0.1$\\%& $1.99\pm0.02$\\
AA$\Omega$&0.68&-22.37&$ 2.04\pm0.02$&$ 1.9\pm0.1$\\%& $2.20\pm0.02$\\
N02E1&$\approx0.1$&-22.68&$1.90\pm0.23$&$6.2\pm2.2$\\%& --\\
N02E2&$\approx0.1$&-22.40&$1.66\pm0.20$&$3.9\pm1.5$\\%& --\\

	\hline
	\hline 
\end{tabular}}
\label{tab:evol}
%\end{minipage}
\end{table}

The stable model and the linear theory (with constant bias) model rise too quickly as the
redshift decreases, excluded at $> 99.99$\% confidence. However, the comoving model also gives 
a good fit to the SDSS/2SLAQ*/AA$\Omega$* data in Fig. \ref{fig:evol}a, as expected from the
lack of evolution shown in Fig. \ref{fig:smallreduce}b. For this model to be exactly correct
it would suggest that there was an inconsistency in these results with the
underlying $\Lambda$CDM halo mass function. More certainly, we conclude
that the evolution of the LRG clustering seems very slow. This general
conclusion agrees with previous work \citep{White07,Wake08}.
The latter author also only found a marginal rejection of the
long-lived model from the large-scale clustering signal (1.8$\sigma$) compared to 1.9$\sigma$ 
here. They found a much stronger rejection of a `passive' evolution model from the 
small-scale LRG clustering and we shall return to this issue in \S \ref{sec:small_evol}.

\subsubsection{LRG dark matter halo masses}
\label{sec:halomass}
The large-scale galaxy bias is roughly the same as that of the dark matter 
haloes which is a known function of mass threshold. Thus by measuring the 
clustering of the LRGs one can infer the typical mass of the haloes they reside 
in. The procedure employed here is similar to that used by \cite{Croom05} and 
\cite{daAngela08} to estimate the dark matter halo masses of QSOs. 

An ellipsoidal collapse model relating a halo bias factor to its mass was 
developed by \cite{Sheth01} as an improvement over an earlier spherical 
collapse model of \cite{Mo96}. In this analysis, we shall use the expression 
given in \cite{Sheth01} and the revised parameters of \cite{Tinker05} which 
were calibrated to give better fits to a wide range of $\sigma_8$ values 
for variants of $\Lambda$CDM model;  

\begin{eqnarray}
b(M_{\mathrm{DMH}},z)&=& 1+\left.\frac{1}{\sqrt{a}\delta_{c}(z)} \right[
\sqrt{a}(a\nu^{2})+\sqrt{a}b(a\nu^{2})^{1-c}\nonumber \\
            & &
~~~~~~-\left.\frac{(a\nu^{2})^{c}}{(a\nu^{2})^{c}+b(1-c)(1-c/2)}\right],
\label{equation:ellipsoidal_bias}
\end{eqnarray}
where $a=0.707$, $b=0.35$ and $c=0.80$. $\nu$ is defined as
$\nu=\delta_{c}(z)/\sigma(M_{\mathrm{DMH}},z)$. $\delta_{c}$ is the critical
density for collapse, and is given by $\delta_{c}=
0.15(12\pi)^{2/3}\Omega_{m}(z)^{0.0055}$ \citep{NFW97}.
The rms fluctuation of the density field as a function of mass
$M_{\mathrm{DMH}}$ at redshift $z$ is
$\sigma(M_{\mathrm{DMH}},z)=\sigma(M_{\mathrm{DMH}})D(z)$ where 
$\sigma(M_{\mathrm{DMH}})$ is given by

\begin{equation}
\sigma(M_{\mathrm{DMH}})^{2} =
\frac{1}{2\pi^{2}}\int_{0}^{\infty}k^{2}P(k)w(kr)^{2}dk
\label{equation:sig_m_int}
\end{equation}
$P(k)$ is the linear power spectrum of density perturbations and $w(kr)$ is the
window function, given by \citep{Peebles80}

\begin{equation}
w(kr)=3\frac{\sin(kr)-kr\cos(kr)}{(kr)^{3}},
\label{equation:w_peebles}
\end{equation}
for a spherical top-hat function. The radius $r$ can be related to mass via

\begin{equation}
r=\left(\frac{3 M_{\mathrm{DMH}}}{4\pi \rho_{0}}\right)^{1/3},
\label{equation:radius_mass}
\end{equation}
where $\rho_{0} = \Omega_{m}^{0}\rho_{crit}^{0}$ is the present mean density of
the Universe, given by $\rho_{0} = 2.78\times 10^{11} \Omega_{m}^{0} h^{2}
M_{\odot}$ Mpc$^{-3}$. Here, we use the transfer function, $T(k)$, fitting
formula given by \citet{Eisenstein98} to construct $P(k)$, assuming our fiducial
cosmology (see \S \ref{sec:large}).
 
The estimated dark matter halo masses of the LRG samples are given in 
Table \ref{tab:evol} and plotted in Fig. \ref{fig:evol}c. 
Note that the formalism of estimating dark matter halo masses from the galaxy
biases used here assumes one galaxy per halo and can overestimate the threshold 
mass for a given value of bias \citep{Zheng07}. This is particularly true when we 
consider the mass estimated from Eq. \ref{equation:ellipsoidal_bias} as the threshold 
mass, minimum mass required for a halo to host at least one galaxy and compare the 
results derived here to $M_{\rm min}$ from the best-fit HOD (\S \ref{sec:HOD}). However, 
if it is used as an estimate for the average mass of the host halo then it is under-estimated 
by $\approx$40 per\,cent compared to the value given by the HOD due to the one galaxy per 
halo assumption.                                                                                                                          

Next, we attempt to fit the derived dark matter halo masses of these LRGs to 
the halo merger framework in hierarchical models of galaxy formation. We use 
the formalism discussed by \cite{Lacey93} to predict the median $M_{\rm{DMH}}$ 
of the descendants of virialised haloes at $z=1$ for a given halo mass and fit 
this to our data. In essence, the model gives the probability distribution of 
the haloes with mass $M_1$ at time $t_1$ evolving into a halo of mass $M_2$ at 
time $t_2$ via merging. Fig. \ref{fig:evol}c shows the best-fit models for the 
$M_{\rm{DMH}}$ evolution estimated in this way. These models appear to be good 
fits to both luminosity bins with the best-fit 
$M_{\rm{DMH}}(z=1)=2.32\pm0.07\times10^{13}\hmsun$ and $1.47\pm0.05\times10^{13}\hmsun$ 
for the $L\ga 3L$* and $\ga 2L$* samples, respectively. 

The most massive haloes hosting these luminous early-type galaxies appear to have 
tripled their masses over the past 7 Gyr (i.e. half cosmic time) in stark contrast 
to the little evolution observed in the LRG stellar masses over the same period 
\citep[see e.g.][]{Wake06,Cool08}. This lack of evolution contradicts the predictions 
in the standard hierarchical models of galaxy formation where one expects the most massive 
galaxies to form late via `dry' merging of many less massive galaxies. However, this comes 
with a caveat that the $M_{\rm{DMH}}$ at $z\sim0$ is an extrapolation 
(assuming \cite{Lacey93} halo merging model) of the $z=0.35-0.7$ measurements and the 
constraint on the $M_{\rm{DMH}}(z=0.1)$ is much weaker than the higher redshift results. 

\subsection{Small-scale clustering evolution} 
\label{sec:small_evol}

Finally, we discuss the evolution of the correlation function at scales
corresponding to $r<1\hmpc$. We concentrate on comparing the
number density matched AA$\Omega$* and 2SLAQ* samples to the SDSS sample. As
can be seen in Fig. \ref{fig:smallreduce}b, while at larger scales the $w(\theta)$ show
amplitudes that are remarkably independent of redshift, at smaller scales
the high redshift AA$\Omega$* sample appears to have a lower amplitude
than the lower redshift surveys. Here, we compare the clustering in non-linear 
regime to two clustering evolution models, namely stable clustering and HOD evolution models.

\subsubsection{Stable clustering model}
\label{sec:stable}
\begin{figure}
	\hspace{-0.7cm}
	\centering
	\includegraphics[scale=0.50]{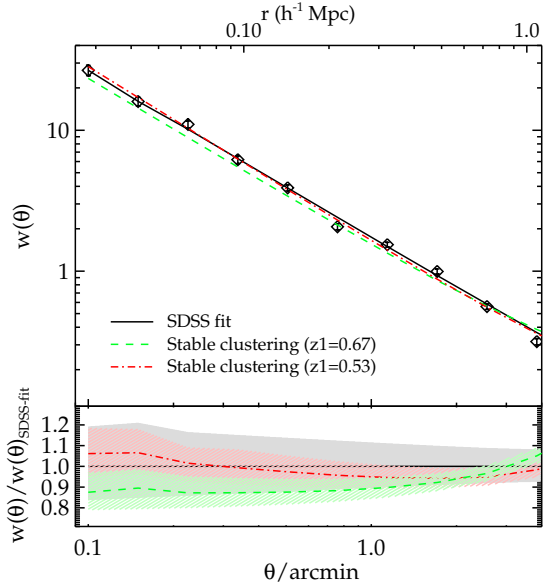}
	\caption{The small-scale $w(\theta)$ at $z=0.35$ evolved from the best-fit 
	double power-law of AA$\Omega$* (green dashed line) and 2SLAQ* (red dot-dashed line) 
	samples, assuming stable clustering model. The ratios of the evolved $w(\theta)$'s to 
	the best-fit double power-law of SDSS sample are shown in the bottom panel. The shaded 
	regions signify $1\sigma$ uncertainties in the best-fit models.}
	\label{fig:evol_small}
	\vspace{-3mm}
\end{figure}

The stable model describes the clustering in the virialised regime and 
hence stable (unchanged) in proper coordinates \citep[e.g.][]{Phillipps78}. 
Therefore, assuming this model one expect the spatial correlation function 
to evolve as $\xi(r)\propto (1+z)^{\gamma-3}$, where $r$ is measured in comoving 
coordinates and $\gamma$ is the power-law slope of the correlation function.  
Fig. \ref{fig:evol_small} shows the small-scale, $r \la 1 \hmpc$, $w(\theta)$ 
of the SDSS sample plus its best-fit double power-law model, comparing to the evolved 
$w(\theta)$ from the $z1=0.53$ and 0.67 best-fit models. Their ratios to the $z=0.35$ 
best-fit model are shown in the bottom panel with the shaded regions represent $1\sigma$ 
uncertainties in the best-fit models. We see that the evolved $z1=0.67$ stable model under-predicts 
the $z=0.35$ $w(\theta)$ somewhat but otherwise is within the $1\sigma$ regions of each other with the 
probability of acceptance $P(<\chi^2)=0.827$. 
The agreement between the evolved $z1=0.53$ and the $z=0.35$ is better, $P(<\chi^2)=0.999$, given that 
the redshift difference is smaller. Note that the stable clustering model over-predicts the clustering amplitude 
at $r\ga1 \hmpc$ which is also observed in Fig. \ref{fig:evol}a as expected.

The physical picture that is suggested is that the inflexion in the
correlation function may represent the boundary between a virialised
regime at small scales and a comoving or passively evolving biased regime
at larger scales. As noted by \cite{Hamilton91} and \cite{Peacock96}, 
the small scale, non-linear, DM clustering is clearly expected from $N$-body 
simulations to follow the evolution of the virialised clustering model. 
However, for galaxies in a $\Lambda$CDM context, the picture may be more 
complicated. 

For example, by comparing the 2SLAQ and SDSS LRG redshift surveys
using the semi-projected correlation function, \cite{Wake08} have
suggested that a passively evolving model is rejected, weakly from the
large scale evolution but more strongly from the evolution at small
scales. \cite{Wake08} interpret the clustering evolution using a HOD
description based on the $\Lambda$CDM halo mass function. Their `passive' model
predicts a far faster evolution at small scales than is given by our stable 
clustering (see Fig. \ref{fig:HODevolve}). Our stable model is certainly passive 
in that it is based on the idea that the comoving number density of galaxies is 
independent of redshift. However, the passive HOD model of \cite{Wake08} requires 
only 7.5 per\,cent of LRGs to merge between z=0.55 and z=0.19 to reconcile the 
slow LRG density and clustering evolution in the $\Lambda$CDM model. We shall 
see in the the next section if this model can also accommodate our z=0.68 
clustering result while maintaining such a low merger rate.   

\subsubsection{HOD evolution}
\label{sec:HODevolution}
In \S \ref{sec:evol_inter}, we found using the large-scale linear bias that 
the long-lived model \citep{Fry96} is only marginally rejected at 1.5-1.9$\sigma$.
This is in good agreement with the similar analysis of \cite{Wake08}. However, 
they argued that if the small-scale clustering signal was also 
taken into consideration, the long-lived model can be ruled out at much higher 
significance ($>$99.9 per\,cent).

Recall that our goodness-of-fit (based on the minimum $\chi^2$) for the halo models is 
rather poor (see table \ref{tab:HODfit}). This may be an indication that a more 
complicated model may be needed, e.g. five-parameters 
HOD and/or a better two halo-exclusion prescription etc., given our small error bars. 
Nevertheless, the HOD fit generally describes the shape and amplitude of our 
measured $w(\theta)$ between 0.1-40 $\hmpc$.
Therefore, at the risk of over-interpreting these HOD fits, we make a further test 
of the long-lived model by evolving the best-fit HODs of the higher redshift samples to 
the SDSS LRG average redshift. 

Following the methods described in \cite[][and references therein]{Wake08}, 
the mean number of galaxies hosted by haloes of mass 
$M$ at later time, $z_0$, is related to the mean number of galaxies in haloes of mass 
$m$, $\left<N(m)\right>$, at earlier time, $z_1$, via

\begin{eqnarray}
\left<N(M)\right>&=&\int_0^M {\rm d}m\,N(m,M)\left<N(m)\right> \nonumber \\ 
                 &=&\int_0^M {\rm d}m\,N(m,M)\left<N_{\rm c}(m)\right>\left[1+\left<N_{\rm s}(m)\right>\right] \nonumber \\
		 &=&C(M)+S(M),
\label{eq:Nmerge}
\end{eqnarray}
where $N(m,M)$ is the conditional halo mass function of \cite{Sheth02} which is the generalisation 
of \cite{Lacey93} results, $C(M)$ and $S(M)$ are the number of objects which used to be central and 
satellite galaxies. 

We then model the central galaxy counts in the low-redshift haloes assuming that the progenitor 
counts in these haloes is `sub-Poisson' \citep{Sheth99,Seo08,Wake08} such that
\begin{equation}
\left<N_{\rm c}(M)\right>=1-\left[1-\frac{C(M)}{N_{\rm max}}\right]^{N_{\rm max}}, 
\end{equation}
where $N_{\rm max}={\rm int}(M/M_{\rm min})$. This model is favoured by the \cite{Wake08} analysis 
and is also seen in the numerical models of \cite{Seo08}. The mean number of satellite galaxies in the 
low-redshift haloes is then given by
\begin{equation}
\left<N_{\rm c}(M)\right>\left<N_{\rm s}(M)\right>=S(M)+f_{\rm no-merge}\left[C(M)-\left<N_{\rm c}(M)\right>\right],
\label{eq:nomerge}
\end{equation}
where $f_{\rm no-merge}$ is the fraction of un-merged low-$z$ satellite galaxies which were high-$z$ 
central galaxies. This model is called `central-central mergers' in \cite{Wake08}, where the more massive 
high-z central galaxies are more likely to merge with one another or the new central galaxy rather than 
satellite-satellite mergers. 

\begin{figure}
\hspace{-0.45cm}
\centering
  \includegraphics[scale=0.53]{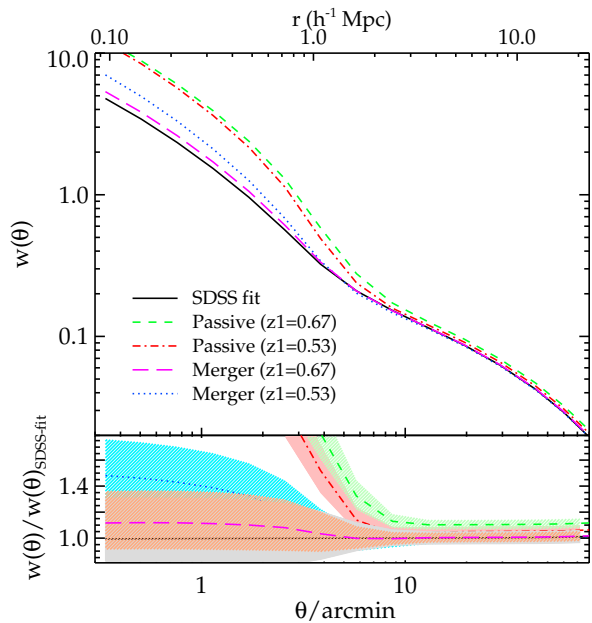}
	\caption{The predicted SDSS LRG $w(\theta)$ from passively ($f_{\rm no-merge}=1$) evolving 
	the best-fit HODs of 2SLAQ* (z1=0.53, red dot-dashed line) and AA$\Omega$* (z1=0.67, green dashed line) 
	samples. The results when central galaxies from high redshift samples are allowed to merge 
	(see text for more detail) are also shown, blue dotted and magenta long-dashed lines. 
	The bottom panel shows the ratios of the evolved $w(\theta)$'s to the SDSS best-fit, the shaded regions 
	signify the $1\sigma$ uncertainties.}
	\label{fig:HODevolve}
\end{figure}

For the long-lived model, we set $f_{\rm no-merge}=1$. The results of passively evolving the best-fit HODs from  
$z1=0.67$ (AA$\Omega$*) and $z1=0.53$ (2SLAQ*) to $z0=0.35$ are shown in Fig. \ref{fig:HODevolve} along with 
the SDSS best-fit model. At large scales ($r \ge 5\hmpc$), the long-lived model can only be marginally rejected 
at no more than $2\sigma$ for the AA$\Omega$* case and is consistent within $1\sigma$ in the case of 2SLAQ*. 
However, if we now consider the small-scale, $r < 1 \hmpc$, clustering signal we see from the bottom panel of Fig. 
\ref{fig:HODevolve} that the long-lived model becomes increasingly inconsistent with the best-fit model at $z=0.35$. 
For $r \ge 0.5 \hmpc$, the long-lived model can be rejected 
at 99.88 and $>99.99$ per\,cent significance using the evolved 2SLAQ* and AA$\Omega$* HODs, respectively. The much 
higher clustering signal at small scales is caused by far too many satellite galaxies in the low-redshift 
haloes being predicted by the long-lived model. This also results in the higher satellite fractions than observed; 
both evolved 2SLAQ* and AA$\Omega$* give $F_{\rm sat}=18\pm1$ per\,cent at $z=0.35$ compared to $8.1\pm1.8$ 
seen in the SDSS best-fit.  

Next, we assume the central-central mergers model \citep{Wake08} and attempt to match the large-scale clustering 
signal of the evolved HOD from high-$z$ to the $z=0.35$ best-fit model. As argued by \cite{Wake08} and here that 
this is more likely to happen than the satellite-satellite merging case. The $f_{\rm no-merge}$ parameters in Eq. 
\ref{eq:nomerge} required to give the best matches to the large-scale clustering amplitude of the SDSS best-fit 
is 0.2 and 0.1 for the 2SLAQ* and AA$\Omega$* case, respectively. The new $w(\theta)$'s determined from these 
models are plotted in Fig. \ref{fig:HODevolve} as the blue dotted and magenta long-dashed lines. We can see that 
the $z1=0.67$ evolved $w(\theta)$ at small scales is in excellent agreement with the SDSS best-fit model. The 
predicted satellite fraction, $F_{\rm sat}=7.8\pm0.9$, is also consistent with the SDSS best-fit value. For the 
$z1=0.53$ case, the small-scale clustering signal is still somewhat stronger that the SDSS best-fit model but 
otherwise are within $1\sigma$ confidence regions of each other, and the predicted $F_{\rm sat}=10.5\pm1.3$ is 
also somewhat higher than the best-fit value. The galaxy number density is reduced due to these central-central 
merger by $\approx$6 and 11 per\,cent for the $z1=0.53$ and 0.67, respectively. However, note that this is 2--3 
times smaller than the fractional errors of our best-fit $n_{\rm g}$, $\approx20$ per\,cent.  

In order to get a handle on the merger rates which can then be compared to the 
previous results of \cite{White07} and \cite{Wake08}, we follow their method of 
adjusting the galaxy number density. This is because for this type of analysis 
the galaxy samples at different redshifts are usually designed to have the same space 
density. Whereas merging means that the space density of the low-$z$ sample must be 
reduced unless there are new galaxies created via merging of the fainter objects which 
fail to be in the high-$z$ sample but become bright enough to be in the low-$z$ sample. 
To account for such an effect by physically removing galaxies in a sample is rather 
difficult to do in practice as argued by \cite{Wake08}. \cite{White07} and 
\cite{Wake08} adjusted the mass-scale of the low-$z$ HOD fit by several per\,cent which 
reduce the space density and increase the clustering signal and hence require lower amount of 
merging of the high-$z$ population needed to match the low-$z$ measurement. Increasing the 
$f_{\rm no-merge}$ factor in Eq. \ref{eq:nomerge} results in a higher galaxy number density and 
clustering signal. Therefore there is only one unique solution of mass-scaling and merging 
fraction that will simultaneously match the galaxy number density and the clustering signal 
(at large scales) of the evolved and best-fit HODs at low-$z$.

\begin{figure}
% \hspace{-0.45cm}
\centering
  \includegraphics[scale=0.5]{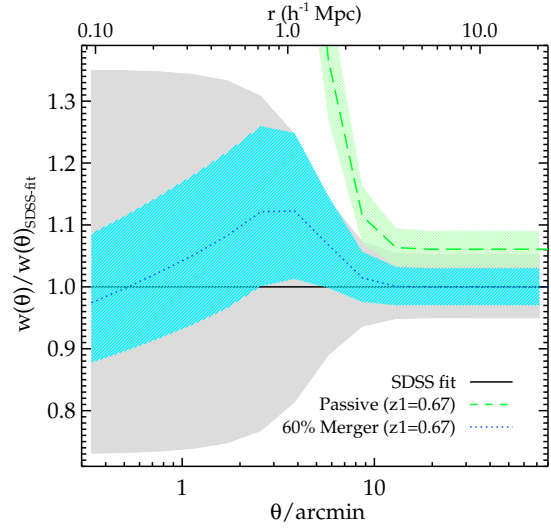}
	\caption{The ratio of the evolved $w(\theta)$ to the SDSS best-fit model with the 
	HOD mass-scale increased by 12 per\,cent.}
	\label{fig:HODmassscale}
\end{figure}

We increase the mass-scale of $z=0.35$ HOD fit by 12 (7) per\,cent 
and allow 60 (50) per\,cent of the $z1=0.67$ (0.53) central galaxies 
to merge in order to get the matched large-scale bias of 2.12 (2.10) and 
$n_{\rm g}=1.12~(1.19) \times 10^{-4} h^3 \Mpc^{-3}$. This yields the merger 
rate between z=0.67 (0.53) and z=0.35 of $\approx6.6$ (5) per\,cent, 
i.e. $\approx2.8$ (3.4) per\,cent Gyr$^{-1}$. The evolved $w(\theta)$ 
divided by the model at $z=0.35$ with increased mass-scaled HOD fit 
is shown in Fig. \ref{fig:HODmassscale}. As noted earlier, the 
reduction in the galaxy number density is small compared to its best-fit 
fractional error which means that our contraints on these merger rates are 
rather weak. However, to first order the merger rates derived here appear to 
be consistent with the value of $2.4\pm0.7$ and 3.4 per\,cent Gyr$^{-1}$ found by 
\cite{Wake08} and \cite{White07}, respectively.

In summary, the combination of the stable clustering and passive evolution model is 
remarkably close to explaining the clustering evolution of the LRGs at small and large scales. 
These models are much simpler than the HOD framework which require an 
understanding of how galaxies populate dark matter haloes and how they and their host 
haloes merge. The galaxy long-lived model in the context of halo framework is significantly 
incompatible with the small-scale clustering data and requires that $\approx2-3$ per\,cent/Gyr 
LRGs to merge in order to explain their slow clustering evolution. 
On the contrary, the stable model requires the comoving number density to be constant 
with redshift. This may suggest that the simple virialised model may only provide a 
phenomenological fit to the small-scale clustering evolution in the context of 
the $\Lambda$CDM model.

% Thus it is interesting that 

\begin{figure*}
	\hspace{-5mm}
	\centering
	\includegraphics[scale=0.47]{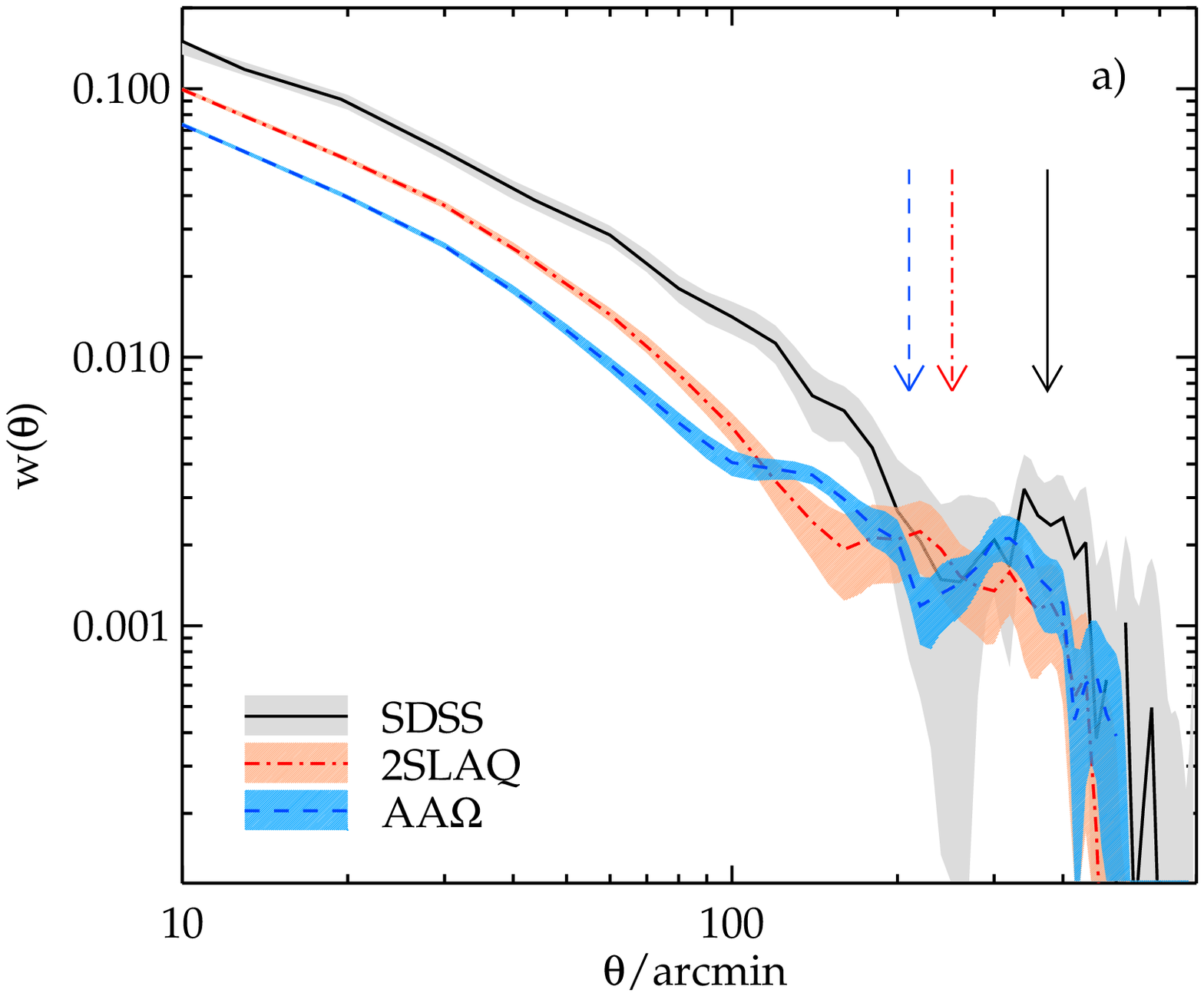}\includegraphics[scale=0.47]{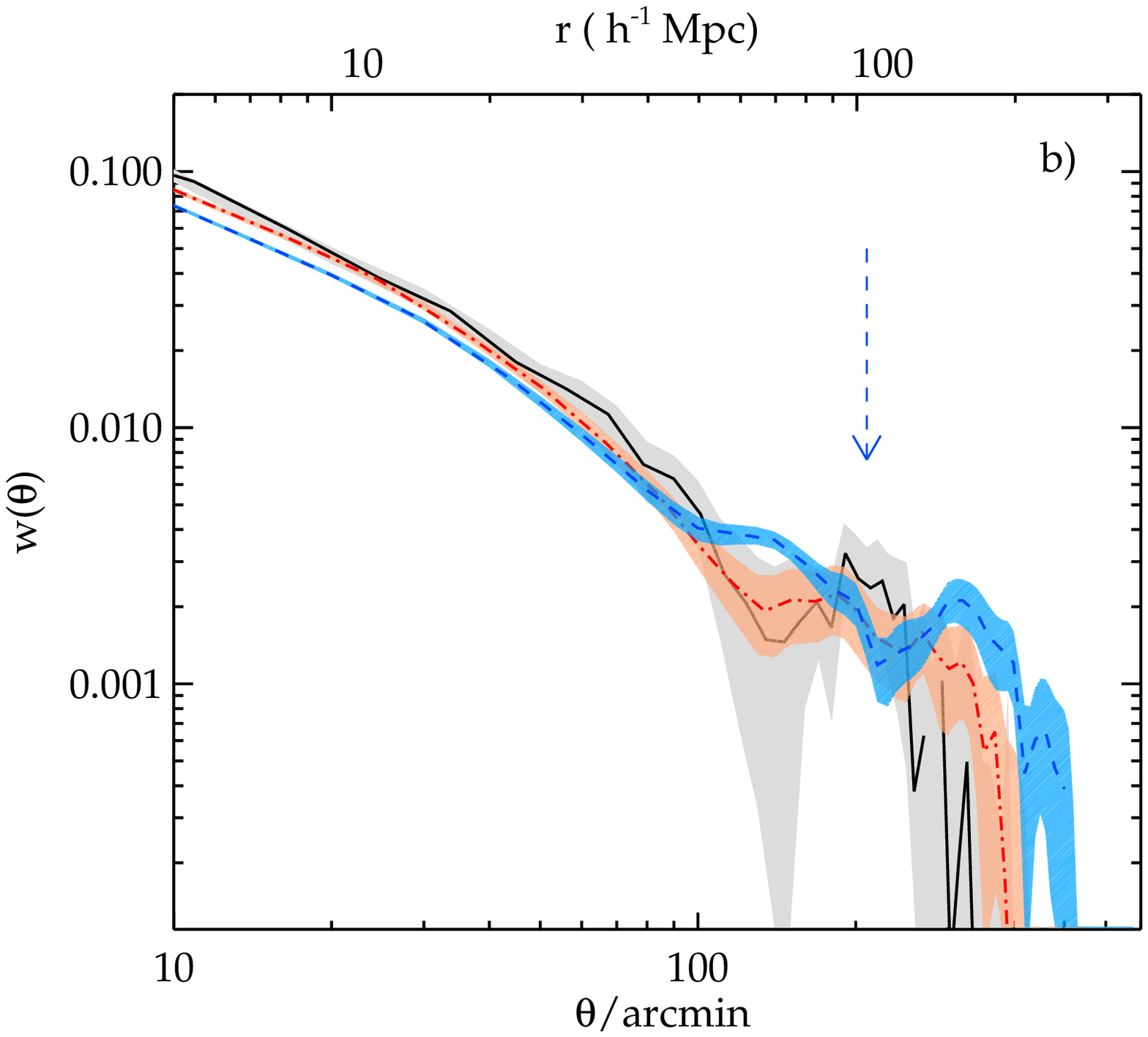}
	\caption{(a): The angular correlation function of the three LRG samples at large
scales. The shaded regions are $1\sigma$ JK errors. The arrow indicates the expected BAO 
angular separation in each sample, assuming our fiducial cosmology. (b): Same as (a) but now 
scaled in the angular direction to the depth of the AA$\Omega$ LRG sample.}

	\label{fig:raw}
\end{figure*}

\begin{figure}
	\hspace{-0.63cm}
	\centering
	\includegraphics[scale=0.47]
	{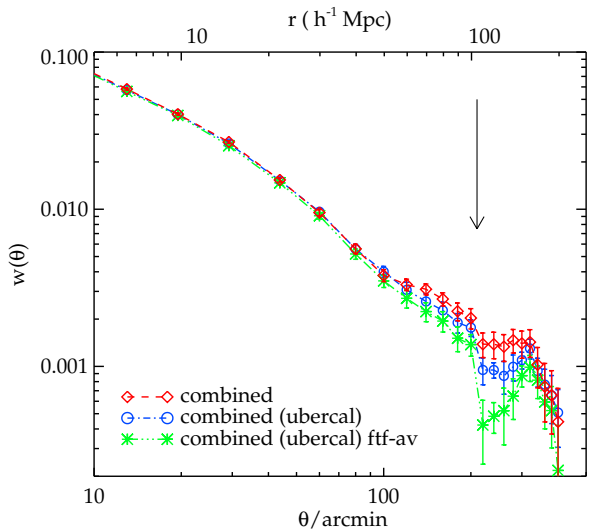}
	\caption{The combined angular correlation function of the three LRG samples 
scaled to the AA$\Omega$ depth, comparing the results when the SDSS standard 
(diamonds) and uber- (circles) calibration are used. Also shown is the average 
field-to-field $w(\theta)$ (asterisks) which represents an attempt to filter out 
any large scale gradients in the SDSS data.}
	\label{combine}
\end{figure}

\section{Searching for the BAO peak}
\label{sec:bao}
Next, we inspect the correlation functions at larger scales to make a search
for the BAO feature. We first present the raw correlation functions in Fig.
\ref{fig:raw}a. Note that the integral constraints (see \S \ref{sec:optimal}) 
are sub-dominant compared to $w(\theta)$'s amplitudes at these scales. 
Each correlation function shows a feature at large scales, 
the most significant detection comes from the AA$\Omega$ sample where the 
clustering signal at $120' < \theta < 500'$ is detected (above zero) at more 
than $4\sigma$ significance, $P(<\chi^2) = 1\times 10^{-6}$ (with covariance 
matrix) and $3.5\sigma$ significance for $200' < \theta < 500'$.  

The question is are these features real or simply due to systematic error? (see \S 
\ref{sec:systematics} for a series of systematic tests). Here, we 
perform a classic scaling test to see if any feature is reproduced at the
different depths of the three LRG samples. 
Given that the samples have intrinsically different $r_0$ (see Table \ref{tab:fit}), 
we choose simply to scale in the angular direction only. The SDSS and 2SLAQ LRG correlation 
functions are scaled in the angular direction to the AA$\Omega$'s depth using the average
radial comoving distance of each sample. 
In Fig. \ref{fig:raw}b, we see that the scaling agreement of the large scale, 
$\theta\approx300'$, features is poor. Although SDSS shows a moderately strong 
peak feature, this is not reproduced at the same comoving physical scale in 
the other two datasets. 

Despite this failure of the scaling test, we now attempt to increase the signal 
to noise ratio by combining the measurements from the three samples using 
inverse quadrature error weighting. Firstly, the SDSS and 2SLAQ 
$w(\theta)$'s are scaled in the angular direction to the depth of the AA$\Omega$
LRGs (radial comoving distance, $\chi\approx 1737 \hmpc$ as opposed to $\approx
1451 \hmpc$ for 2SLAQ and $\approx 970 \hmpc$ for SDSS) where their amplitudes and errors
are then interpolated to the AA$\Omega$'s angular bins (i.e. Fig. \ref{fig:raw}b). 
The amplitudes of the scaled SDSS and 2SLAQ $w(\theta)$'s are then normalised to that 
of the AA$\Omega$ sample's at $10'$. This involves lowering SDSS and 2SLAQ amplitudes 
by 25 and 15 per\,cent, respectively. 
The resulting correlation function is presented in Fig.
\ref{combine} with the arrow showing the expected position of the BAO peak. Note 
that due to the relatively small statistical errors of the
AA$\Omega$ LRG compared to other samples, the $w(\theta)$ result is dominated by
the AA$\Omega$ sample, therefore the possible SDSS peak at $\approx 100 \hmpc$ is
not evident in the combined sample. There also seems to be an indication of 
an excess out to possibly $200 \hmpc$ (see \S \ref{sec:systematics} for a 
robustness test of this excess clustering signal).

\begin{figure*}
% 	\hspace{-0.5cm}
	\centering
\includegraphics[width=17cm]{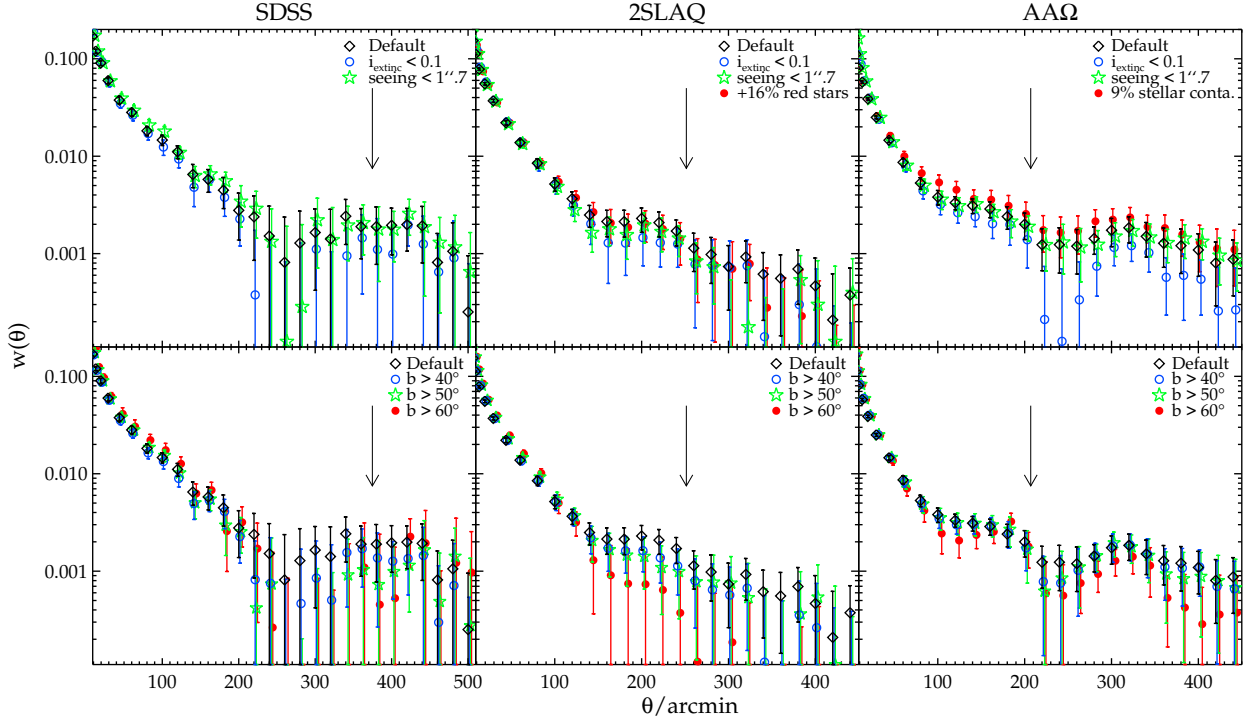}
	\caption{The angular correlation functions for SDSS, 2SLAQ and AA$\Omega$ samples (left to right), 
	measured with varying dust extinction limit, astronomical seeing and different star-galaxy 
	separation (top row), compared to our `Default' results. Also shown is the effect of low galactic latitude 
	region exclusion for each sample (bottom row). Note that for the $b>60\deg$, the sample size is reduced by 
	60 per\,cent. In each case, an arrow indicates the expected position of the BAO peak assuming our 
	fiducial cosmology.}
	\label{fig:sys}
\end{figure*}

Using the ubercalibration \citep{Padmanabhan08} instead of the standard
calibration, we find similar results at small and intermediate scales but somewhat 
lower amplitude at $\approx100 \hmpc$ although the results
agree within the $1\sigma$ error (see Fig. \ref{combine}). This means the
correlation functions at small and intermediate scales including the parameters
derived (e.g. power-law fits, linear biases, dark matter halo masses) in the
earlier parts are not affected by which calibration we use. The biggest
difference, although less than $1.5\sigma$, is observed at scales larger 
than $120 \hmpc$ and up to $150 \hmpc$ where the correlation signal is small 
and hence more prone to possible systematics. 
The weak dependence of $w(\theta)$ at very large scales 
on the different calibrations may be an indication that this apparent extra peak at 
$\theta\approx300'$ could indeed be a systematics effect. We shall return to this in 
\S \ref{sec:systematics}.
      
We also tested whether the $200\hmpc$ excess can be eliminated by taking
the average $w(\theta)$ from $15\times20 $ deg$^2$ subfields. 
The result, after integral constraint correction, is shown in
Fig. \ref{combine}. The $200\hmpc$ excess persists even though there is some change at
smaller scales. Given the model dependence introduced by the integral
constraint correction (Eq. \ref{equ:w_IC}), hereafter, we shall use the correlation function of
the ubercal sample measured using our normal method.

\subsection{Testing for systematic effects}
\label{sec:systematics}
We have performed a series of tests to check our results against possible 
systematic effects. The tests include exclusions of high dust extinction and 
`poor' astronomical seeing regions, an improved star-galaxy separation for 
the AA$\Omega$ sample and effects of possible contamination by clustered stars. 

First, we exclude the regions where the $i$-band extinction is greater than 0.1 mag 
which discards $\approx 20$ per\,cent of the data. The results are shown in the top row of 
Fig. \ref{fig:sys}. For 2SLAQ and AA$\Omega$ samples, the results appear to be lower than 
the main measurements but otherwise remain within $1\sigma$ statistical errors of each other. 
Although the amplitudes at $\theta \ge 220'$ are somewhat lower than the default AA$\Omega$ result, 
the excess at $\theta \ga 300'$ still persists. 
We then investigate the effect of excluding the regions with `poor' astronomical seeing, 
the limit of $1''.7$ is used following the SDSS `poor' seeing definition which discards 
$\approx 30$ per\,cent of the data. The results here are in good agreement with the main 
results with the exception of a few angular bins around $320'$ of the 2SLAQ sample where 
they are somewhat (non-significantly) lower than the default measurements.

Next, we attempt to reduce the stellar contamination fraction in the AA$\Omega$ sample. 
As a reminder, our default (optimised) star--galaxy separation algorithm 
(see \S \ref{sec:data}) leaves $\approx 16$ per\,cent stellar contamination in the sample 
while losing genuine LRGs only at a sub-per\,cent level. Here, we impose a more aggressive 
star--galaxy separation cut which reduces the contamination level to $\approx$9 per\,cent
at the expense of nearly halving the number of genuine AA$\Omega$ LRGs. The cut is a combination 
of the fitted `de Vaucouleurs' radius as a function of $z_{\rm deV}$ magnitude and the correlation 
between the `de Vaucouleurs' and fiber magnitudes in $z$-band. The $w(\theta)$ measurement for 
this new AA$\Omega$ sample after correction by a factor of $1/(1-f)^2$, where $f=0.09$ 
is shown in the top-right panel of Fig. \ref{fig:sys}. This is in good agreement 
with the main results. 

We test our earlier asssumption (\S \ref{sec:method}, see also \citealt{Blake08}) that the effect of 
the stellar contamination is simply a dilution of $\delta_{\rm g}$ by $(1-f)$ and hence the amplitude 
of galaxy--galaxy correlation function by $(1-f)^2$, where $f$ is the contamination fraction. We add 
a sample of red stars to the 2SLAQ sample at the 16 per\,cent level, similar to what we expect in 
the AA$\Omega$ sample. The stars are selected from SDSS photometric objects which are classified as 
`star'. The colour selections have been matched to that of the AA$\Omega$-LRG sample. The sample is then 
randomly selected to have the number of objects at 16 per\,cent of the 2SLAQ sample. They therefore follow 
the stellar distribution with galactic latitude. 
The $w(\theta)$ result after correction by $1/(1-f)^2$ is shown in the 
top-middle panel of Fig. \ref{fig:sys} and is found to be in excellent agreement with the main 
2SLAQ result. We do not see any evidence of a slope change which may arise from a possible clustering 
of the stellar contaminants at large scales, at least for the contamination level expected in our sample. 

We apply various minimum galactic latitude cuts on the data in order to test for any systematic error.
Such systematics (if they exist) could be due to the gradient caused by galactic dust extinction and/or 
different stellar contamination fractions which one might expect to be worse in the lower galactic 
latitude regions. Note that in our default datasets $\approx 95$ per\,cent of the data are 
at $b \ge 30 \deg$. The results of applying the galactic latitude cuts of $b \ge 40 \deg$, $50 \deg$ and 
$60 \deg$ are shown in the bottom row of Fig. \ref{fig:sys}. Note that with the $b \ge 60 \deg$ limit, 
$\approx 60$ per\,cent of the data are discarded. 
The 2SLAQ results appear to be marginally dependent on the galactic latitude limits. 
In the AA$\Omega$ sample the results are in good agreement with the main measurement although 
the $b \ge 60 \deg$ limit appears to be $\approx 1\sigma$ lower in some angular bins. 

\begin{figure}
% 	\hspace{-0.5cm}
	\centering
\includegraphics[scale=0.5]{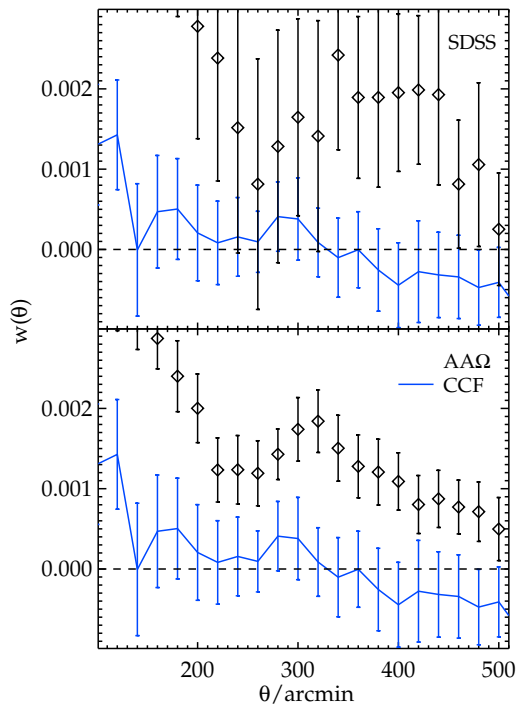}
	\caption{The auto correlation functions for SDSS (top panel) and AA$\Omega$ (bottom panel) samples, comparing to 
	the CCF between the two samples (blue solid lines).}
	\label{fig:CCF}
\end{figure}

Finally, we cross-correlate the SDSS and AA$\Omega$ samples. The redshift distributions of the two samples 
are well separated with only slight overlap (see Fig.\ref{nz}). Therefore any residual signal in their 
cross-correlation function, CCF, at large scales can be used as an evidence for systematic errors. 
The CCF is shown in Fig. \ref{fig:CCF}, comparing to the auto correlation functions of the SDSS and 
AA$\Omega$ samples in the top and bottom panels, respectively. The CCF has much lower signal than the 
ACF at $\theta < 120'$ and is consistent with zero, $P(<\chi^2) = 0.997$, between $120' < \theta < 500'$ 
whereas the AA$\Omega$ $w(\theta)$ signal is detected at more than $4\sigma$ significance (see above) in 
the same angular ranges.

We note that \cite{Ross11} have suggested that there is a systematic effect associated with 
the area effectively masked by foreground stars which may be important in terms of a systematic that 
may produce excess clustering at large scales. However, such an effect would predict a decrease 
in galaxy density at low galactic latitudes and this is not seen in our samples 
\citep[see Fig. 9 of][]{Sawangwit10}. If anything, the opposite effect is seen in our data with 
an increase in density towards lower galactic latitude which may be caused by stellar contamination. 
Here, we have tested our $w(\theta)$ measurements by successively cutting out data at low galactic 
latitudes. Although the 2SLAQ results may show some marginal dependence on the galactic latitude cut, 
the AA$\Omega$ results seem reasonably unaffected (see Fig. \ref{fig:sys}(f)). This may be due to the higher 
stellar contamination fraction in AA$\Omega$ sample which means that the effect seen by \cite{Ross11} 
may not be directly applicable to the AA$\Omega$ sample. 

We conclude that the apparent clustering excess at $\approx 300'$ in the AA$\Omega$ sample appears 
to be reasonably robust against most of the systematic tests we performed here. However, one might argue 
that the weakening of the excess signal when $i_{\rm extinc} > 0.1$ regions ($\approx$20 per\,cent) are 
excluded and the marginal dependence on the galactic latitude cuts of the 2SLAQ results 
may be taken as some evidence for systematic effects. On the other hand, the SDSS-AA$\Omega$ cross-correlation 
test also tends to limit the size of possible systematic errors.
  
\subsection{Model comparisons}
\label{sec:compare_mod}
\subsubsection{Standard $\Lambda$CDM model}
\label{sec:bao_LCDM}
First, we compare the measured angular correlation function to the perturbation
theory prediction in the standard $\Lambda$CDM Universe. To compute the theoretical
prediction, we proceed in the same manner as described in \S \ref{sec:large},
calculating $w(\theta)$ by projecting $\xi(r)$ which is a Fourier transform
of a non-linear $P(k)$. However, here we assume the best-fit cosmological 
parameters from \cite{Eisenstein05}, a flat $\Lambda$CDM model with $\Omega_m h^2=0.13$, 
$\Omega_b h^2=0.024$, $h=0.7$ and $n=0.98$. And unlike in \S \ref{sec:large}, 
the non-linear modelling of the BAO peak using only HALOFIT is not 
adequate. The BAO peak in the correlation function can also be broadened (and perhaps slightly shifted) 
by the non-linear gravity suppression of the higher harmonics in the power spectrum via mode coupling 
\citep{Meiksin99}. To model such an effect, we follow \cite{Eisenstein05} and smoothly interpolate between 
the linear power spectrum and the `no-wiggle' spectrum with the same overall 
shape but with the acoustic oscillations erased. This is done mathematically by

\begin{figure*}
	\centering
        \hspace{-0.5cm}
       
\includegraphics[scale=0.6]{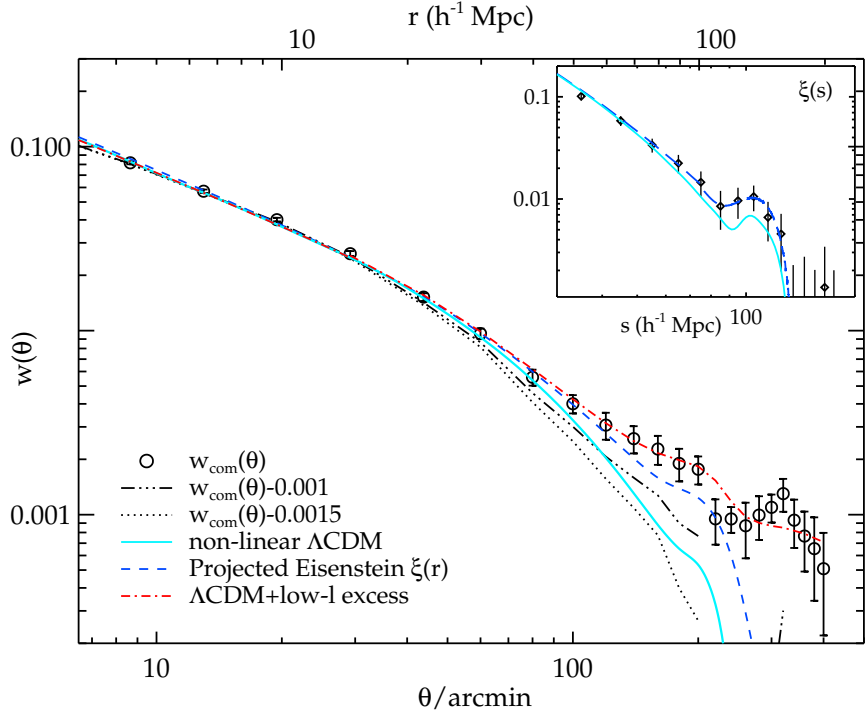}
        \caption{The combined $w(\theta)$ (open circles) compared to the projections 
of non-linear $\Lambda$CDM model plus mode coupling, scale-dependent redshift-space 
distortion and halo bias correction (cyan solid line) and the \citet{Eisenstein05} 
$\xi(s)$ (blue dashed line). The red dot-dashed line is the $\Lambda$CDM model plus 
low-$l$ power excess (\S \ref{sec:excess}). The dash-dot-dotted and dotted line 
shows the effect of subtracting the data by 0.001 and 0.0015, respectively. 
The $\xi(r)$ models used in the $w(\theta)$ projection are given as an inset 
together with the \citet{Eisenstein05} measurement (diamonds). Here, the same 
symbols are used for the \citet{Eisenstein05} and non-linear $\Lambda$CDM 
$\xi(s)$ models as for the $w(\theta)$ models above.}
	\label{fig:combine1}
\end{figure*}

\begin{equation}
P(k)=P_{\rm{lin}}\left[x+\frac{T_{\rm{nw}}(k)\times (1-x)}{T_{\rm{lin}}(k)}\right]^2,
\end{equation}
where $P_{\rm{lin}}$ is linear matter power spectrum, $T_{\rm{nw}}(k)$ and $T_{\rm{lin}}(k)$ are `no-wiggle' 
and linear transfer functions computed from \cite{Eisenstein98} and $x=\exp(-k^2a^2)$ with $a=7\hmpc$ chosen 
to fit the BAO suppression seen in their N-body simulations. 

The $P(k)$ is then corrected for non-linear gravitational collapse using the HALOFIT software. 
The final $P(k)$ is then transformed to $\xi(r)$ using Eq. \ref{eq:pkxirtransform}. Although 
the scale-dependent redshift-distortion and halo bias correction is 
weak at these scales, we follow \cite{Eisenstein05} and multiply the correlation function by the square of 
$1+0.06/[1+(0.06r)^6]$ (solid line in their Fig. 5), again chosen to fit what is seen in the N-body simulations. 
Such a correction is small at the BAO scale, only sub-percent at $r\ga25\hmpc$ and increases to $\approx10$\% at 
$10\hmpc$. We then correct for the linear redshift-space distortion, the $\xi(s)$ amplitude is enhanced
relative to the real-space correlation function, $\xi(r)$, such that \citep{Kaiser87}
\begin{equation}
\xi(s)=\left(1+\frac{2\beta}{3}+\frac{\beta^2}{5}\right)\xi(r).
\label{equ:kaiser}
\end{equation}
Here, we assume $\beta=0.45$ for these LRG samples (see \citealt{Ross07}). 
The final $\xi(s)$ model prediction with galaxy bias $b=2.09$ for SDSS-LRG 
(see \S \ref{sec:large}) is shown (cyan solid line) in the inset of Fig. 
\ref{fig:combine1}. \cite{Eisenstein05} find this model to be 
a good fit to their $\xi(s)$ data with the best-fit $\chi^2=16.1$ on 
17 degrees of freedom for a particular set of cosmological parameters given above. 
We then computed $w(\theta)$ from the $\xi(r)$ derived above via Eq. \ref{mylim}. 
Although the model (cyan solid line in Fig. \ref{fig:combine1}) was found to be consistent with 
the LRG $\xi(s)$, it is inconsistent with our $w(\theta)$ measurement, especially 
at $r\ga 60 \hmpc$ or $\theta \ga 120'$. 
The uber-cal AA$\Omega$ $w(\theta)$ between $40'-400'$ (corresponding to $20 \la r \la 200 \hmpc$) 
are incompatible with the model at 99.8 per\,cent level ($\chi^2$=39.3 over 18-1 d.o.f 
with covariance matrix). We note that this rejection may be associated with the apparent 
clustering excess at $\theta \ga 200'$, which still could be subject to systematics.
 
Next, we compare our $w(\theta)$ to the best estimate of $\xi(s)$ at the BAO
scale as measured by \cite{Eisenstein05}. Although these measurements
may have been superseded by DR7 spectrosopic LRG clustering analyses based on
larger samples, these more recent estimates are usually in reasonable
agreement with the results of \cite{Eisenstein05}, whether they are in
correlation function \citep{Martinez09,Kazin10} or power spectrum
\citep{Percival10} form. For our comparison, we thus simply make a
polynomial fit to the best estimate $\xi(s)$ of \cite{Eisenstein05}
(blue dashed line in the inset of Fig. \ref{fig:combine1}). 
The polynomial-fit $\xi(s)$ is Kaiser de-boosted (Eq. \ref{equ:kaiser}) 
to give $\xi(r)$ by assuming $\beta=0.45$. The $\xi(r)$ is then corrected for 
the linear growth between $z=0.35$ and $z=0.68$ which reduces the amplitude by 
$\approx30$\%. The resulting model has similar amplitude with the expected 
AA$\Omega$-LRG $\xi(r)$ because the SDSS and AA$\Omega$-LRG linear biases 
are coincidentally the same (see \S \ref{sec:large}). The model is then projected 
to $w(\theta)$ using Eq. \ref{mylim} and is shown as a blue dashed line in Fig. 
\ref{fig:combine1}. Our result appears to be in good agreement with the model 
up to $\approx 120 \hmpc$ given statistical uncertainties in our measurement and the 
$\xi(s)$ data. Beyond $\approx 120 \hmpc$, our $w(\theta)$ shows a higher clustering 
amplitude as noted above. 

Summarising, the $w_{\rm com}$ result appears consistent with the $w(\theta)$
prediction based on the \cite{Eisenstein05} best estimate of $\xi(s)$ (at least out to 
$\approx 120 \hmpc$) but not with the prediction based on the flat $\Lambda$CDM model 
due to the apparent large-scale clustering excess in the $w(\theta)$. 
This means that given the size of error bars of the \cite{Eisenstein05} result, the $\Lambda$CDM 
model is quite compatible with the $\xi(s)$ data but given the much smaller statistical 
error on $w(\theta)$, in this case our measurements are inconsistent with the $\Lambda$CDM model. 
While the feature observed at $\approx 300'$ persists in most of the systematic tests we performed 
on the AA$\Omega$ samples (\S \ref{sec:systematics}), a few of these tests, e.g. dust extinction, 
indicate there is still the possibility that systematic errors are affecting the 
$w(\theta)$. Therefore, if we now assume that the excess signal 
at $\approx150 \hmpc$ is an indication of a systematic and subtract 0.001 
to 0.0015, the level of the excess amplitude at this point in $w_{\rm com}$ 
(see Fig. \ref{fig:combine1}), we obtain the $w(\theta)$ results as shown by 
the dash-dot-dotted and dotted lines. These two lines now bracket the flat 
$\Lambda$CDM result. Thus the issue of the disagreement between the
$w(\theta)$ result and the $\Lambda$CDM model seems to rest on the reality of 
the apparent clustering excess at large scales.

\subsubsection{low-$\ell$ power excess?}
\label{sec:excess}
Recently, \cite{Thomas10} \citep[see also][]{Padmanabhan07,Blake07,Thomas11}
has also found a significant excess in their angular power 
spectrum, $C_\ell$, at the low multipoles relative to the best-fit $\Lambda$CDM 
models. They used photometric-redshift catalogues of the LRGs at $z\approx0.5$ 
similar to our 2SLAQ sample. The most significant ($\approx 4\sigma$) 
low-$\ell$ power excess is observed in the highest redshift bin. 
The author carried out various systematic checks and found no indication 
of such an effect. 
While the clustering excess only affects $C_\ell$ at multipoles smaller than the 
acoustic oscillations in Fourier space, unfortunately in configuration space the 
effect is expected on a wider range of scales and could affect our $w(\theta)$ 
BAO measurement. 

\begin{figure}
        \hspace{-7mm}
	\centering
        \includegraphics[scale=0.5]{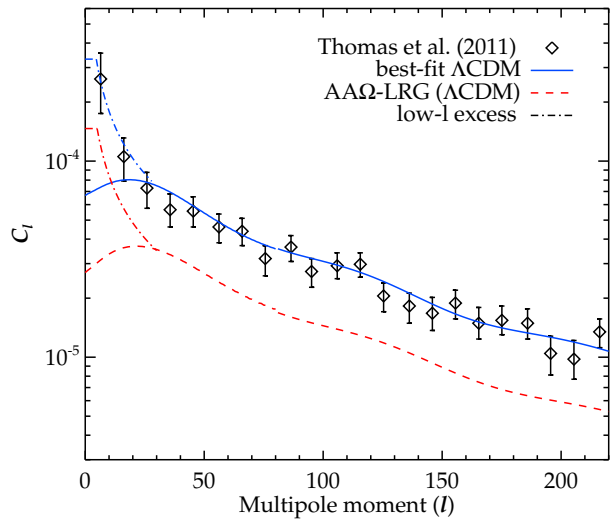}
	\caption{The angular power spectrum of the $0.6 < z < 0.65$ MegaZ-DR7 LRG \citep{Thomas11} 
	with significant power excess at low multipoles (diamonds). The low-$\ell$ power excess plus 
	the best-fit $\Lambda$CDM model of the $0.6 < z < 0.65$ $C_\ell$ (blue dot-dashed and solid lines) 
	is extrapolated to the AA$\Omega$ redshift range (red dot-dashed and dashed lines).} 
        \label{fig:cl_excess}
\end{figure}

Fig. \ref{fig:cl_excess} shows the \cite{Thomas10} $C_\ell$ and the excess power model 
used to predict the AA$\Omega$ $w(\theta)$. For further detail of the low-$\ell$ power 
excess model, see \cite{SawangwitThesis}. 
The resulting $w(\theta)$ model with the amplitude normalised to fit the data at $\theta=40'-400'$ is shown 
as the red dot-dashed line in Fig. \ref{fig:combine1}. The model appears to be consistent with 
our $w_{\rm com}$ (13 per\,cent confidence level, $\chi^2=23.6$ over $18-1$ d.o.f) at $r\approx 20-200 \hmpc$. 
Therefore we note that the excess clustering signal in our $w(\theta)$ is in good agreement with that observed in 
the $C_\ell$ by \cite{Thomas10}. The fact that the excess power in the $C_\ell$ taken the form of an $\ell \approx10$ spike, 
suggests that this excess in $w(\theta)$ is due to something other than acoustic oscillations in the power 
spectrum.

We note that evidence for a large-scale ($>150 \hmpc$) correlation 
function excess has also been detected in the NVSS radio source survey 
by \cite{Blake02} and \cite{Xia10}. We have compared our results with theirs
and find that our correlation function shows a similar shape but a
factor of 2-3$\times$ lower amplitude. If the excess clustering signal observed here is 
real then it could be evidence for non-Gaussianity \citep{Xia10} or for the
gauge dependence of the matter power spectrum on the largest scales 
\citep{Lin01,Yoo09}. But until this feature is detected in an independent galaxy dataset, 
there will always be the possibility that it is caused by some unknown systematics. 
Certainly, if the $\Lambda$CDM model were correct then we would have to conclude that 
this excess was caused by systematics at the level of $\Delta w\approx0.001-0.0015$ 
in the photometric AA$\Omega$-LRG sample.

\section{Summary and Conclusions}
\label{sec:conclusion}

We have presented here a new and detailed analysis of the angular correlation 
function of the Luminous Red Galaxies extracted from the SDSS DR5 photometric 
catalogue. All the necessary information for inferring the spatial clustering 
is obtained and calibrated through redshift surveys of sample subsets. Our
conclusions are as follows;

\begin{enumerate}
\item We measured the angular correlation function of the LRGs at three
different redshifts, namely 0.35, 0.55 and 0.68 and found the results to be well
aproximated by power-laws at small and intermediate scales.
\item With the large samples in terms of the numbers of objects and volume
cover by the data, we see the deviation from the canonical single power-law at
high significance.
\item The data are better fitted by a double power-law where the large-scale 
( $\ga 1-2 \hmpc$) slope is equal to that of the conventional single power-law,
i.e. $\gamma\approx1.8$.
\item The form of the angular correlation functions at large scales are consistent 
with the expectation of the linear perturbation theory in the flat standard 
$\Lambda$CDM Universe.
\item The LRG linear bias is high, $b_g \approx 2.0$, as expected for massive
luminous early-type galaxies and the clustering strength is found to be strongly
linked to the sample intrinsic brightness.
\item The best-fit HOD models suggest that these LRGs reside in the masssive 
dark matter haloes, $10^{13}-10^{14} h^{-1} M_{\odot}$, and are typically 
central galaxies in their dark matter halo hosts, with the satellite fraction 
no more than 10 per\,cent.
\item The clustering evolution at intermediate scales ($1 < r < 20 \hmpc$) is 
remarkably slow and may be approximately explained by a long-lived model or even 
a no--evolution model. The long-lived model may be in line with the observed
passive evolution of the LRG luminosity function, consistent with a constant 
comoving LRG space density in this redshift range. This latter conclusion would also
apply in the case that the no--evolution (comoving) model were found to fit
better but in this case the observations may require a significantly higher bias.
\item Using the \cite{Lacey93} framework, our $M_{\rm DMH}(z)$ measurements are well
fitted by the model where halo mass is grown via merging of progenitors with
masses of $\approx 1.4 \times 10^{13} \hmsun$ and $\approx 2.3 \times 10^{13}
\hmsun$ from $z=1$, for haloes that typically host $L\ge2L$* and $\ge3L$*
galaxies, respectively. We found that these dark matter haloes have tripled their
masses over the last half of cosmic time (although see the caveat given at the end 
of \S \ref{sec:halomass}) whereas it has been claimed that the LRG stellar masses 
have grown by less than 50 per\,cent \citep{Cool08}.
\item At small scales ($r < 1 \hmpc$) the clustering evolution appears slightly 
faster at fixed luminosity and the clustering increases towards lower redshift, 
consistent with a virialised clustering model. 
Since our virialised model assumes a constant comoving LRG space density, a combination of
this stable clustering model at small scales and the long--lived model at
intermediate scales could be consistent with the idea that merging of LRGs may
not change the LRG space density significantly out to $z\approx0.7$.
\item However, the evolution based on HOD and the $\Lambda$CDM halo merging framework requires 
that $\sim 2-3$ per\,cent/Gyr of the LRGs merge with each other in order to explain the 
small-scale clustering evolution, consistent with the results of \cite{White07} and \cite{Wake08}.

\item In our AA$\Omega$-LRG result we find a BAO 
peak at a level consistent with the best estimate of $\xi(s)$ obtained   
by Eisenstein et al (2005). But, given the small size of our statistical
errors, these results deviate significantly, $\approx 4\sigma$, from the 
standard $\Lambda$CDM prediction because of an apparent large-scale 
clustering excess.

\item The excess clustering signal generally persists after a series of systematic tests we performed. 
However, a few of these tests did change the feature somewhat, suggesting that it could still be 
caused by some unknown systematic effects.    

\item If the $\Lambda$CDM model were correct then we would have to conclude that 
this excess was caused by systematics at the level of $\Delta w\approx0.001-0.0015$ 
in the photometric AA$\Omega$-LRG sample.

\item Otherwise, the excess signal in our $w(\theta)$ relative to the standard
$\Lambda$CDM model appears to be in good agreement with the $C_\ell$ power excess
at low $l$ observed by other authors who used photo--z LRG samples at $z\approx0.5$.

\item If real, the large-scale clustering excess may be interpreted as an evidence 
for a non-standard cosmological model, e.g. primordial non-gaussianity or general 
relativistic effects. However, more, independent, data is required to check the 
reality of this clustering excess. 
\end{enumerate}

\section*{Acknowledgements}
US acknowledges financial support from the Institute for the 
Promotion of Teaching Science and Technology (IPST) of The Royal 
Thai Government. We thank Michael J.\,I. Brown for useful comments. 
We thank the referee for useful comments which have improved the 
quality of the paper. 
We also thank all the present and former staff of the 
Anglo--Australian Observatory for their work in building and 
operating the 2dF and AAOmega facility. 

Funding for the SDSS and SDSS-II has been provided by the Alfred
P. Sloan Foundation, the Participating Institutions, the National
Science Foundation, the U.S. Department of Energy, the National
Aeronautics and Space Administration, the Japanese Monbukagakusho, the
Max Planck Society, and the Higher Education Funding Council for
England. The SDSS Web Site is {\tt http://www.sdss.org/}.

The SDSS is managed by the Astrophysical Research Consortium for the
Participating Institutions. The Participating Institutions are the
American Museum of Natural History, Astrophysical Institute Potsdam,
University of Basel, Cambridge University, Case Western Reserve
University, University of Chicago, Drexel University, Fermilab, the
Institute for Advanced Study, the Japan Participation Group, Johns
Hopkins University, the Joint Institute for Nuclear Astrophysics, the
Kavli Institute for Particle Astrophysics and Cosmology, the Korean
Scientist Group, the Chinese Academy of Sciences (LAMOST), Los Alamos
National Laboratory, the Max-Planck-Institute for Astronomy (MPIA),
the Max-Planck-Institute for Astrophysics (MPA), New Mexico State
University, Ohio State University, University of Pittsburgh,
University of Portsmouth, Princeton University, the United States
Naval Observatory, and the University of Washington.

\bibliographystyle{mn2e}
\bibliography{wtheta}

\appendix
\section{Angular Correlation Functions and Covariance Matrices}
\label{appen:wtheta}
\begin{figure*}
% \begin{minipage}{5cm}
	\centering
%          \vspace{-50mm}
	\includegraphics[scale=0.4]{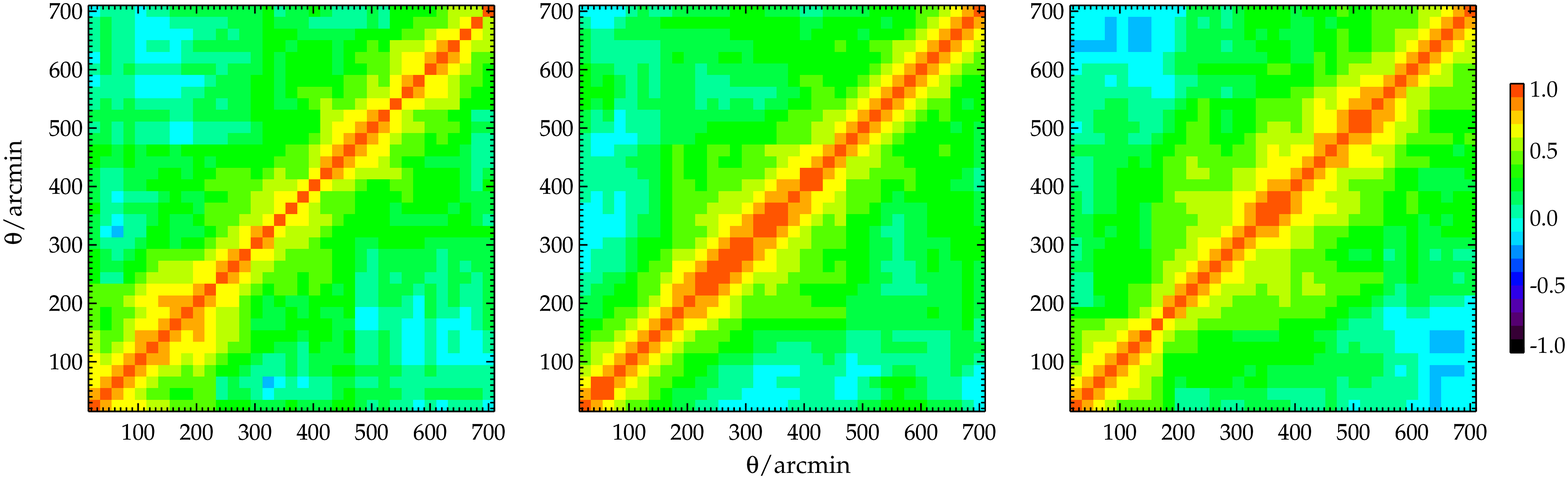}
	\caption{The correlation coefficients, $\mathbfss{r}_{ij}$, out to very large angular separations. 
	These are derived from the covariance matrices (Eq. \ref{eq:corre_coeff}) via 96 jackknife re-sampling 
	fields. Three panels show $\mathbfss{r}_{ij}$ for SDSS, 2SLAQ and AA$\Omega$ -LRG samples from left to right.}    
	\label{fig:corre_large}
% \end{minipage}
\end{figure*}

At the referee's request, we tabulate the angular 
correlation functions (Table \ref{tab:wtheta}) measured 
from the three photometric LRG samples studied in this paper. 
The full covariance matrices in the form of correlation 
coefficients are also shown in Fig. \ref{fig:corre_large}.

\begin{table}

	\caption{The measured angular correlation functions for 
	the SDSS, 2SLAQ and AA$\Omega$-LRG and their $1\sigma$ JK errors.}
%\begin{minipage}{5cm}
\hspace{-0.6cm}
%\centering
% \resizebox{!}{2.0cm}
	{\begin{tabular}{lccc}

	\hline
        \hline
$\theta (')$&    SDSS                      &            2SLAQ                 &       AA$\Omega$                \\
      \hline
    0.100  &  $26.78 \pm   2.37$ &   $9.85 \pm      0.39$ &    $6.27  \pm    0.24 $ \\
    0.150  &  $15.96 \pm   1.47$ &   $7.40 \pm      0.14$ &    $4.65  \pm    0.10 $ \\
    0.225  &  $11.09 \pm   0.56$ &   $4.54 \pm      0.085$ &   $2.95  \pm    0.057 $ \\
    0.337  &  $6.10 \pm   0.33$ &    $2.95 \pm      0.050$ &   $1.86  \pm    0.033 $ \\
    0.506  &  $3.93 \pm   0.19$ &    $1.83 \pm      0.026$ &   $1.11  \pm    0.016 $ \\
    0.759  &  $2.04 \pm   0.090$ &   $1.09 \pm      0.020$ &   $0.65  \pm    0.014 $ \\
    1.139  &  $1.55 \pm   0.061$ &   $0.68 \pm      0.011$ &   $0.419  \pm    0.0095 $ \\
    1.708  &  $1.00 \pm   0.038$ &   $0.416 \pm      0.0057$ & $0.282  \pm    0.0059 $ \\
    2.562  &  $0.56 \pm   0.025$ &   $0.285 \pm      0.0061$ & $0.213  \pm    0.0036 $ \\
    3.844  &  $0.31 \pm   0.019$ &   $0.199 \pm      0.0038$ & $0.151  \pm    0.0023 $ \\
    5.766  &  $0.22 \pm   0.012$ &   $0.152 \pm      0.0026$ & $0.112  \pm    0.0020 $ \\
    8.649  &  $0.171 \pm  0.0081$ &   $0.113 \pm      0.0019$ &   $0.083  \pm    0.0013 $ \\
    12.97 &  $0.118 \pm   0.0053$ &   $0.078 \pm      0.0018$ &   $0.057  \pm    0.0011 $ \\
    19.46 &  $0.091 \pm   0.0055$ &   $0.055 \pm      0.0012$ &   $0.0405  \pm    0.00077 $ \\
    29.19 &  $0.060 \pm   0.0041$ &   $0.038 \pm      0.0011$ &   $0.0264  \pm    0.00062 $ \\
    43.78 &  $0.038 \pm   0.0031$ &   $0.0226 \pm      0.0009$ &  $0.0157  \pm    0.00060 $ \\
    60.00 &  $0.028 \pm   0.0023$ &   $0.0144 \pm      0.0008$ &  $0.0093  \pm    0.00053 $ \\
    80.00 &  $0.018 \pm   0.0020$ &   $0.0086 \pm      0.00076$ & $0.0056  \pm    0.00051 $ \\
    100.0  &  $0.014 \pm   0.0019$ &  $0.0054 \pm      0.00067$ & $0.0040  \pm    0.00045 $ \\
    120.0  &  $0.011 \pm   0.0017$ &  $0.0034 \pm      0.00060$ & $0.0039  \pm    0.00036 $ \\
    140.0  &  $0.0071 \pm   0.0018$ &   $0.0024 \pm      0.00061$ &   $0.0035  \pm    0.00027 $ \\
    160.0  &  $0.0063 \pm   0.0014$ &   $0.0019 \pm      0.00064$ &   $0.0029  \pm    0.00032 $ \\
    180.0  &  $0.0045 \pm   0.0013$ &   $0.0021 \pm      0.00065$ &   $0.0024  \pm    0.00039 $ \\
    200.0  &  $0.0026 \pm   0.0014$ &   $0.0020 \pm      0.00060$ &   $0.0020  \pm    0.00039 $ \\
    220.0  &  $0.0020 \pm   0.0014$ &   $0.0022 \pm      0.00062$ &   $0.0011  \pm    0.00035 $ \\
    240.0  &  $0.0014 \pm   0.0013$ &   $0.0019 \pm      0.00058$ &   $0.0014  \pm    0.00039 $ \\
    260.0  &  $0.0014 \pm   0.0015$ &   $0.0015 \pm      0.00045$ &   $0.0015  \pm    0.00040 $ \\
    280.0  &  $0.0017 \pm   0.0011$ &   $0.0013 \pm      0.00044$ &   $0.0018  \pm    0.00032 $ \\
    300.0  &  $0.0020 \pm   0.00077$ &  $0.0013 \pm      0.00045$ &   $0.0021  \pm    0.00038 $ \\
    320.0  &  $0.0016 \pm   0.00091$ &  $0.0015 \pm      0.00045$ &   $0.0021  \pm    0.00043 $ \\
    340.0  &  $0.0032 \pm   0.0010$ &   $0.0013 \pm      0.00053$ &   $0.0019  \pm    0.00048 $ \\
    360.0  &  $0.0025 \pm   0.0010$ &   $0.0011 \pm      0.00047$ &   $0.0016  \pm    0.00048 $ \\
    380.0  &  $0.0023 \pm   0.0011$ &   $0.0012 \pm      0.00045$ &   $0.0016  \pm    0.00045 $ \\
    400.0  &  $0.0025 \pm   0.0010$ &   $0.0010 \pm      0.00045$ &   $0.0013  \pm    0.00041 $ \\
    420.0  &  $0.0017 \pm   0.0011$ &   $0.00054 \pm      0.00045$ &   $0.0007  \pm    0.00041 $ \\
    440.0  &  $0.0020 \pm   0.0012$ &   $0.00064 \pm      0.00042$ &   $0.0006  \pm    0.00038 $ \\
    460.0  &  $0.0003 \pm   0.0012$ &   $0.00017 \pm      0.00045$ &   $0.0008  \pm    0.00038 $ \\
    480.0  &  $0.0006 \pm   0.0014$ &   $0.00002 \pm      0.00047$ &   $0.0005  \pm    0.00039 $ \\
    500.0  &  $-0.0001\pm   0.0012$ &   $0.00018 \pm      0.00051$ &   $0.0005  \pm    0.00044 $ \\

	\hline
	\hline 
\end{tabular}}
\label{tab:wtheta}
%\end{minipage}
\end{table}

%+Bibliography
%\begin{thebibliography}{99}
%\bibitem{Label1} ...
%\bibitem{Label2} ...
%\end{thebibliography}
%-Bibliography

%\bibitem[\protect\citeauthoryear{{Adelman-McCarthy} \& et
%  al.}{{Adelman-McCarthy} et~al.}{2006}]{Adelman06}
%{Adelman-McCarthy} J.~K.,  et al. 2006, \apjs, 162, 38

\label{lastpage}
\end{document}